\documentclass[aps,pra,onecolumn,amsmath,amssymb,showpacs, superscriptaddress,notitlepage,nofootinbib]{revtex4-2}

\usepackage{graphicx}

\usepackage{amsmath,amssymb,mathtools,physics}
\usepackage[utf8]{inputenc}

\usepackage[dvipsnames]{xcolor}

\usepackage{hyperref}
\hypersetup{
	colorlinks   = true, 
	urlcolor     = blue, 
	linkcolor    = blue, 
	citecolor   = blue 
}

\newcommand\scalemath[2]{\scalebox{#1}{\mbox{\ensuremath{\displaystyle #2}}}} 

\usepackage[capitalize]{cleveref}

\usepackage [english]{babel}
\usepackage [autostyle, english = american]{csquotes} \MakeOuterQuote{"}

\newcommand{\affvigocenter}{Vigo Quantum Communication Center, University of Vigo, Vigo E-36315, Spain}
\newcommand{\affuvigo}{Escuela de Ingeniería de Telecomunicación, Department of Signal Theory and Communications, University of Vigo, Vigo E-36310, Spain}
\newcommand{\affatlantic}{atlanTTic Research Center, University of Vigo, Vigo E-36310, Spain}
\newcommand{\afftoyama}{Faculty of Engineering, University of Toyama, Gofuku 3190, Toyama 930-8555, Japan}
\newcommand{\affwaterloo}{Institute for Quantum Computing and Department of Physics and Astronomy, University of Waterloo, Waterloo, Ontario N2L 3G1, Canada}

\begin{document}
	
	\title{Security of quantum key distribution with imperfect phase randomisation}
	\author{Guillermo Currás-Lorenzo}
	\affiliation{\affvigocenter} \affiliation{\affuvigo} \affiliation{\affatlantic} 
	\affiliation{\afftoyama}
	\author{{Shlok Nahar}}
	\affiliation{\affwaterloo}
	\author{{Norbert L\"utkenhaus}}
	\affiliation{\affwaterloo}
	\author{Kiyoshi Tamaki}
	\affiliation{\afftoyama} 
	\author{Marcos Curty}
	\affiliation{\affvigocenter} \affiliation{\affuvigo} \affiliation{\affatlantic}
	
	\begin{abstract}
		The performance of quantum key distribution (QKD) is severely limited by multiphoton emissions, due to the photon-number-splitting attack. The most efficient solution, the decoy-state method, requires that the phases of all transmitted pulses are independent and uniformly random. In practice, however, these phases are often correlated, especially in high-speed systems, which opens a security loophole. Here, we address this pressing problem by providing a security proof for decoy-state QKD with correlated phases that offers key rates close to the ideal scenario. Our work paves the way towards high-performance secure QKD with practical laser sources, and may have applications beyond QKD.
	\end{abstract}
	
	\maketitle
	
	\section{Introduction}
	
	Quantum key distribution (QKD) allows two users, Alice and Bob, to securely establish a symmetric cryptographic key over an untrusted channel controlled by an adversary, Eve, with unlimited computational power\;\cite{loSecureQuantum2014,xuSecureQuantum2020}. The security of QKD is based on information theory and the laws of quantum mechanics. However, a practical implementation of a QKD protocol is only secure if it meets all the assumptions made in its corresponding security proof. For example, the early proofs\;\cite{shorSimpleProof2000,mayersQuantumKey1996} of the widely-known BB84 protocol\;\cite{bennettQuantumCryptography1984} assumed the availability of single-photon sources, which are difficult to achieve in practice. Instead, implementations of the protocol typically rely on laser sources that emit weak coherent pulses (WCPs), either with or without randomised phases, which are vulnerable to the photon-number-splitting attack\;\cite{brassardLimitationsPractical2000}  and to an unambiguous state discrimination attack\;\cite{dusekUnambiguousState2000}, respectively. This has a severe impact on the obtainable secret-key rate and limits the maximum distance to a few tens of kilometers\;\cite{gottesmanSecurityQuantum2004,loSecurityQuantum2007}. 
	
	The most efficient solution to this problem is known as the decoy-state method\;\cite{hwangQuantumKey2003,loDecoyState2005,wangBeatingPhotonNumberSplitting2005,maPracticalDecoy2005}, and is currently used by the majority of commercial QKD systems. It requires the users to emit phase-randomised (PR) WCPs of various intensities, and exploits the fact that PR-WCPs are diagonal in the Fock basis, with each photon-number component containing no information about the intensity it originated from. Thanks to this, one can use the observed detection statistics to characterize the effect of the channel on different photon-number states, and derive tight bounds on the fraction of the sifted key that originates from single-photon emissions, as well as on its phase-error rate. As a result, one can ideally obtain a secret-key rate comparable to that offered by single-photon sources\;\cite{limConciseSecurity2014}. 
	
	However, generating \textit{perfect} PR-WCPs, i.e.\ WCPs whose phase is uniformly and independently random in $[0,2\pi)$, may be challenging in certain scenarios, particularly at high repetition rates. The {most} common approach {to randomise the pulse phase} is to operate the laser under gain-switching conditions\;\cite{yuanUnconditionallySecure2007,dixonGigahertzDecoy2008,liuDecoystateQuantum2010,lucamariniEfficientDecoystate2013,boaronSecureQuantum2018}, i.e.\ {to} turn the laser on and off between pulses. However, due to the difficulty in attenuating the intracavity field {of the laser} strongly enough to ensure significant phase diffusion, experiments suffer from residue correlations between the phases of consecutive pulses\;\cite{kobayashiEvaluationPhase2014,grunenfelderPerformanceSecurity2020}, which invalidate the standard decoy-state analysis. {As an alternative, one can also actively randomise the phase {of each emitted pulse} by using a random number generator and a phase modulator \cite{zhaoExperimentalQuantum2007}, and security proofs have been proposed to deal with the resulting discretisation effect \cite{caoDiscretephaserandomizedCoherent2015,curras-lorenzoTwinFieldQuantum2021}. However, due to memory effects in the phase modulator {and the electronics that control it}\;\cite{grunenfelderPerformanceSecurity2020}, this approach may also suffer from correlations, which the existing proofs do not take into account.
		
	Because of this discrepancy between the existing security proofs of decoy-state QKD and its practical implementations, the security of the latter} is not sufficiently guaranteed, which is an important {open} problem in the field.
 	{Here}, we {address this problem} by proving {its} security in the presence of phase correlations between consecutive pulses{, which arise when running gain-switched laser sources at high repetition rates.} Importantly, our simulation results suggest that decoy-state QKD is robust against this imperfection, and that one could obtain key rates close to the ideal scenario when using currently-available {high-speed} laser sources.

 	\section{Assumptions and Protocol description}

 	{Clearly, the secret key rate obtainable in the presence of imperfect phase randomisation should depend on the strength of the imperfection. The case in which the phases are not random is known to result in a very poor performance \cite{loSecurityQuantum2007}, while one may expect that, if the source emits a train of pulses whose phases are close to the ideal scenario (i.e., all being independent and uniformly distributed), one should also be able to obtain a performance that is close to ideal. Thus, determining the obtainable key rate inevitably requires a certain degree of source characterisation, with the only question being which specific parameters are the relevant ones. Our proof demonstrates that only two parameters need to be characterized. The main parameter that determines the protocol performance, which we denote as the source quality $q \in (0,1]$, evaluates how close each individual phase is to being uniformly random from the perspective of an eavesdropper that holds all possible side-information about it, i.e., that has knowledge of all previous and following phases that are correlated with it. 
 	
 	The other relevant parameter in our security proof is the correlation length $l_c$, which does not affect the asymptotic key rate obtainable, but does have consequences in the post-processing step, see \cref{subsec:protocol_description} below. We remark that the case $l_c=0$ --- i.e., the case in which the phases are independent but not uniformly random,} which may be relevant if Eve performs an active laser seeding attack \cite{sunEffectSource2015} --- {has already been considered in Refs.~\cite{naharDecoyStateQuantum2022,naharImperfectPhaseRandomisation2023a}; our proof becomes similar to that of these works for this scenario.} {For concreteness, in this work,} we focus on the applicability of our proof to the case of naturally-occurring phase correlations.
 	
 	 	\subsection{Assumptions {of our proof}}
 	
 	{The sequence of phases $\Phi_1 ... \Phi_N$ of Alice's pulse train constitute a discrete-time stochastic process whose joint distribution can be represented} by a probability density function (PDF) $f(\phi_1...\phi_N)$. Our proof does not require a precise characterisation of this distribution; it requires just two pieces of knowledge, which we state as the following two assumptions:
	
{	(A1) The stochastic process $\Phi_1 ... \Phi_N$  has at most $l_c$ rounds of memory, for some finite and known $l_c$. That is, for all rounds $i$,
	\begin{equation}
		\label{eq:assumption_lc}
		f(\phi_i \vert \phi_{i-1}... \phi_{1})=f(\phi_i \vert \phi_{i-1}... \phi_{i-l_c}).
	\end{equation}
	%

	
	(A2) The conditional PDF of $\Phi_i$ given all other phases is lower bounded, i.e., for all $i$ and some known $0<q\leq1$,}
	\begin{equation}
		\label{eq:assumption_q}
		f(\phi_i \vert \phi_{1}... \phi_{i-1} \phi_{i+1}...\phi_{N}) = f(\phi_i \vert \phi_{i-l_c}... \phi_{i-1} \phi_{i+1}...\phi_{i+l_c})  \geq \frac{q}{2\pi}.
	\end{equation}
{The equality in \cref{eq:assumption_q} follows from \cref{eq:assumption_lc}.}

	
	{In addition, for simplicity, we consider that the phase randomisation process is not affected by the intensity modulation and bit-and-basis encoding processes; and for concreteness, we also consider that neither the latter processes nor Bob's measurement setup suffer from imperfections. Precisely, we assume that} (A3) Alice's choice of intensities and the phase of her pulses are independent; (A4)
	Alice's (bit-and-basis) encoding operations commute with the process that (imperfectly) randomises the phase of her pulses; (A5) Alice's choice of bit, basis and intensity for round $i$ only affects the $i\textrm{-th}$ pulse; (A6) Alice's encoding operations are characterised and identical for all rounds; (A7) the intensities of Alice's pulses perfectly match her choices; and (A8) the efficiency of Bob's measurement is independent of his basis choice. We note that  previous works have {investigated} the security of QKD when some of {these} assumptions are not met\;{\cite{wangGeneralTheory2008,pereiraQuantumKey2020,zapateroSecurityQuantum2021,sixtoSecurityDecoystate2022,tamakiDecoystateQuantum2016,wangFinitekeySecurity2018,wangMeasurementdeviceindependentQuantum2021,mizutaniQuantumKey2019,yoshinoQuantumKey2018,pereiraQuantumKey2019,navarretePracticalQuantum2021,navarreteImprovedFiniteKey2022,fung2009security,zhangSecurityProof2021,jiangRobustTwinfield2023}}.

	\subsection{Protocol description}
	\label{subsec:protocol_description}
	
	(1) {For} each round, Alice probabilistically selects a random intensity $\mu$ from a predetermined set and attempts to generate a PR-WCP of that intensity. Then, she selects a random bit $b$ and basis $\omega \in \{Z,X\}$, and applies an encoding operation $\hat{V}_{b_\omega}$  
	to her pulse, satisfying $\hat{V}_{b_\omega}^{\dagger}\hat{V}_{b_\omega}= \mathbb{ I}$\footnote{We note that, in our description, $\hat{V}_{b_\omega}$ is an isometric operator, {but not necessarily} a unitary, which reflects the fact that the space after the encoding operation {can be (and in fact typically is)} larger than the original space. For example, in BB84 polarization encoding, ${\{}\hat{V}_{b_\omega}{\}}$ encodes a single mode of light in a prefixed input polarization into four possible outcome polarizations whose creation operators can be expressed as a {linear} function of those of horizontal and vertical polarizations. {To compute the numerical results in \cref{fig:graph}, we have assumed that ${\{}\hat{V}_{b_\omega}{\}}$ are ideal BB84 $Z$- and $X$-basis encoding operators, such that, for any input coherent state $\ket{\alpha}$,} {$\hat V_{0_Z}\ket{\alpha} = \ket{\alpha} \ket{0}$, $\hat V_{1_Z}\ket{\alpha} = \ket{0} \ket{\alpha}$, $\hat V_{0_X}\ket{\alpha} = \ket*{\frac{\alpha}{\sqrt 2}}\ket*{\frac{\alpha}{\sqrt 2}}$ and $\hat V_{1_X}\ket{\alpha} = \ket*{\frac{\alpha}{\sqrt 2}}\ket*{-\frac{\alpha}{\sqrt 2}}$}. However, we remark that our analysis is valid for any set of characterized encoding operators, regardless of the dimension of the encoding space.}.
	
	In the security proof, we consider the following equivalent process for the state preparation: (1a) Alice generates  {$\ket{\sqrt{\nu}}^{\otimes N}$, where $\nu \geq \mu\,\, \forall \mu$;} (1b) she applies an imperfect phase randomisation operation to the pulse train{, obtaining
	\begin{equation}
		\rho_{\rm laser} =  \int_{0}^{2 \pi} d \phi_1 ...  \int_{0}^{2 \pi} d \phi_N f(\phi_1 ...\phi_N) \hat P(\ket{\sqrt{\nu} e^{i \phi_1}}) \otimes ... \otimes \hat P(\ket{\sqrt{\nu} e^{i \phi_N}}),
	\end{equation}
		 {where $\hat P (\ket{\mkern 1mu\cdot\mkern 1mu}) = \ketbra{\mkern 1mu \cdot\mkern 1mu}$;}} (1c) she probabilistically selects all the intensities $\mu_1,...,\mu_N$  and attenuates each pulse to match her selection; and (1d) she probabilistically makes all bit and basis choices $b_{\omega_1},...,b_{\omega_N}$, and applies $\hat{V}_{b_{\omega_1}}...\hat{V}_{b_{\omega_N}}$ to her pulse train. Note that, because of Assumptions (A3) and (A4), steps (1b), (1c) and (1d) commute. 
	
	(2) For each incoming signal, Bob chooses a random basis $Z$ or $X$, and measures the incoming pulse. 
	
	(3) Bob announces which rounds were detected and, for these rounds, both Alice and Bob reveal their basis choices, and Alice reveals her intensity choices. They define their sifted keys as the bit outcomes of the detected rounds in which both chose the $Z$ basis and Alice chose a certain signal intensity $\mu_s$. Also, they define the test rounds as the detected rounds in which Bob used the $X$ basis, and reveal their bit values for these rounds. Moreover, they assign each round $i$ to a group $w \in \{0,...,l_c\}$ according to the value $w = i \bmod (l_c+1)$. The $w\textrm{-th}$ sifted subkey is defined as the fraction of the sifted key belonging to group $w$.
	
	(4) Alice and Bob sacrifice a small fraction of the $w\textrm{-th}$ sifted subkey to estimate its bit-error rate, and use the detection statistics of the $w\textrm{-group}$ test rounds to estimate its phase-error rate. Then, they perform error correction and privacy amplification independently for each subkey.
	
	\section{Security proof}

	The {main idea and contribution} of our security proof is {finding} an equivalence between the actual scenario described above, in which Alice's source is correlated and partially uncharacterised, and an alternative scenario in which, within the $w$-group rounds, Alice prepares characterised and uncorrelated state{s} that {are} close to a PR-WCP, and then applies a global quantum operation that imprints the correlations present in the actual source{, }which, from the perspective of the security proof, can be considered to be part of the Eve-controlled quantum channel. In this alternative scenario, 
	it is straightforward to prove the security of the $w\textrm{-th}$ subkey using numerical techniques; by doing so, we also indirectly prove the security of the $w\textrm{-th}$ subkey in the actual protocol. By repeating this procedure for all $w \in \{0,...,l_c\}$, we can independently prove the security of each subkey, and guarantee the security of the concatenated final key due to the universal composability {property} of each individual security proof. For more information on this latter argument, we refer the reader to Appendix\;C of Ref.\;\cite{mizutaniSecurityRoundrobin2021}, as well as to Ref.\;\cite{yoshinoQuantumKey2018} for an example of its application in the case $l_c =1$.
	
	\subsection{Reduction to the ($w\textrm{-th}$) alternative scenario}

	 {Let $\mathcal{G}_w\;(\mathcal{G}_{\overline w})$ be the set of rounds that belong (do not belong) to group $w$, let $\vec{\phi}_{\mathcal{G}_{w}}$ ($\vec{\phi}_{\mathcal{G}_{\overline w}}$) be a particular joint value for all phases in $\mathcal{G}_{w}$ ($\mathcal{G}_{\overline w}$), let $f(\vec{\phi}_{\mathcal{G}_{\overline w}})$ be the joint marginal PDF of the phases in $\mathcal{G}_{\overline w}$, and let $f(\vec{\phi}_{\mathcal{G}_{w}}\vert \vec{\phi}_{\mathcal{G}_{\overline w}})$ be the joint conditional PDF of the phases in $\mathcal{G}_{w}$ given $\vec{\phi}_{\mathcal{G}_{\overline w}}$.}

	After the {chain} of equivalences {(E1)-(E4)} {below}, the actual protocol is reduced to the $w\textrm{-th}$ alternative scenario{, in which Alice's source is characterised and uncorrelated within the rounds in $\mathcal{G}_{w}$. For the first equivalence, note that{, due to Assumption (A1),} the phases in $\mathcal{G}_w$ are conditionally independent of each other given knowledge of the phases in $\mathcal{G}_{\overline w}$, i.e.,
	\begin{equation}
		\label{eq:cond_indep}
		f(\vec{\phi}_{\mathcal{G}_{w}}\vert \vec{\phi}_{\mathcal{G}_{\overline w}}) = \prod_{i \in \mathcal{G}_w} f(\phi_i \vert \vec{\phi}_{\mathcal{G}_{\overline w}}),
	\end{equation}
	{as shown in \cref{app:proof_of_independence}.}
	
	(E1) Let us assume that Alice performs step (1b) in the following way. First, she chooses $\vec{\phi}_{\mathcal{G}_{\overline w}}$ according to the marginal PDF $f(\vec{\phi}_{\mathcal{G}_{\overline w}})$. Then, for each round $i$, (a) if $i\in \mathcal{G}_{\overline w}$, she shifts the phase {of the pulse} by her selected fixed value $\phi_i \in \vec{\phi}_{\mathcal{G}_{\overline w}}$; (b) if $i\in \mathcal{G}_w$, she shifts the phase according to the conditional PDF $f(\phi_i \vert \vec{\phi}_{\mathcal{G}_{\overline w}})$. 
	
	Conditioned on a specific value $\vec{\phi}_{\mathcal{G}_{\overline w}}$, the state generated by Alice is
	\begin{equation}
		\scalemath{0.9}{\rho_{\vec{\phi}_{\mathcal{G}_{\overline w}}} = \bigotimes_{i' \in \mathcal{G}_{\overline w}} \hat P(\ket{\sqrt{\nu} e^{i \phi_{i'}}}) \bigotimes_{i \in \mathcal{G}_{w}} \int_{0}^{2 \pi} d \phi_i f(\phi_i \vert \vec{\phi}_{\mathcal{G}_{\overline w}})  \hat P(\ket{\sqrt{\nu} e^{i \phi_{i}}}),}
	\end{equation}
	and due to \cref{eq:cond_indep}, the overall generated state is
		\begin{equation}
		\int_{0}^{2\pi}\!\!\! ... \int_{0}^{2\pi}  d \vec{\phi}_{\mathcal{G}_{\overline w}} f(\vec{\phi}_{\mathcal{G}_{\overline w}})  \rho_{\vec{\phi}_{\mathcal{G}_{\overline w}}} = \rho_{\rm laser}.
	\end{equation}

	For the next equivalence, note that Alice could attenuate her pulses before applying the phase shifts above, rather than afterwards. Also, for all $i \in \mathcal{G}_w$, 
	\begin{equation}
	\label{eq:condition_w_group}
	f(\phi_i \vert \vec{\phi}_{\mathcal{G}_{\overline w}}) \\ \geq \frac{q}{2\pi},
\end{equation}
due to Assumptions\;(A1) and (A2). As a consequence, instead of shifting the $i$-th phase according to the PDF $f(\phi_i \vert \vec{\phi}_{\mathcal{G}_{\overline w}})$ {when $i \in \mathcal{G}_w$}, Alice could have equivalently done the following {\cite{naharDecoyStateQuantum2022,naharImperfectPhaseRandomisation2023a}}: {to} flip a biased coin $C_i$ such that $C_i = 0$ with probability $q$, and (a) if $C_i=0$, shift the phase by a uniformly random value, (b) if $C_i=1$, shift it according to the PDF}
\begin{equation}
	\label{eq:condition_w_group2}
	f(\phi_i \vert \vec{\phi}_{\mathcal{G}_{\overline w}}, C_i=1) = \frac{f(\phi_i \vert \vec{\phi}_{\mathcal{G}_{\overline w}}) - \frac{q}{2 \pi}}{1-q}.
\end{equation}
The equivalence is due to
\begin{equation}
	\begin{gathered}
		\label{eq:dist_C_1}
			\scalemath{0.95}{f(\phi_i \vert \vec{\phi}_{\mathcal{G}_{\overline w}}) = q f(\phi_i \vert \vec{\phi}_{\mathcal{G}_{\overline w}}, C_i=0) + (1-q) f(\phi_i \vert \vec{\phi}_{\mathcal{G}_{\overline w}}, C_i=1),}
	\end{gathered}
\end{equation}
{where $f(\phi_i \vert \vec{\phi}_{\mathcal{G}_{\overline w}}, C_i=0) = 1/2\pi$.}

(E2) Instead of steps (1a) to (1c), for each round $i$, Alice probabilistically chooses an intensity $\mu$, and (a) if $i \in \mathcal{G}_{\overline w}$, Alice prepares $\ket*{\sqrt\mu}$; (b) if $i \in \mathcal{G}_w$, Alice prepares 
\begin{equation}
	\label{eq:rho_model_def}
	\rho_{\rm model}^{\mu} \coloneqq q \, \rho_{\rm PR}^{\mu} + (1-q) \ketbra{\sqrt{\mu}},
\end{equation}
where $\rho_{\rm PR}^{\mu}$ is a perfect PR-WCP {of intensity $\mu$}. Then, Alice chooses $\vec{\phi}_{\mathcal{G}_{\overline w}}$ according to the PDF $f(\vec{\phi}_{\mathcal{G}_{\overline w}})$ and, for each round $i$, (a) if $i\in \mathcal{G}_{\overline w}$, {she shifts the phase by her selected fixed value} $\phi_i \in \vec{\phi}_{\mathcal{G}_{\overline w}}$; (b) if $i\in \mathcal{G}_w$, {she shifts} the phase according to the PDF $f(\phi_i \vert \vec{\phi}_{\mathcal{G}_{\overline w}}, C_i=1)$ in \cref{eq:condition_w_group2}.

Clearly, the rounds in $\mathcal{G}_{\overline w}$ are identical in both (E1) and (E2). The rounds in $\mathcal{G}_w$ are also identical. Alice's phase shift does not affect the $\rho_{\rm PR}^{\mu}$ term in \cref{eq:rho_model_def}, and it causes the $\ketbra*{\sqrt{\mu}}$ term to {acquire} the phase distribution in \cref{eq:condition_w_group2}. Thus, the overall phase distribution of the pulse after the shift is $f(\phi_i \vert \vec{\phi}_{\mathcal{G}_{\overline w}})$, due to \cref{eq:dist_C_1}. 
{We can represent Alice's probabilistic selection of $\vec{\phi}_{\mathcal{G}_{\overline w}}$ together with all the phase shifts described above as a single global quantum operation $\mathcal{E}_w$.}

	(E3) Same as (E2), but Alice applies her encoding operations $\hat{V}_{b_{\omega_1}}...\hat{V}_{b_{\omega_N}}$ before $\mathcal{E}_w$, rather than afterwards, which is possible thanks to Assumption\;(A4).
	
	(E4) Since $\mathcal{E}_w$ is now the last operation before the quantum channel, we consider that Alice does not {actually} apply it. Eve may or may not apply $\mathcal{E}_w$ as part of her attack, putting her in a position that is never less advantageous than in the previous scenarios. Thus, if the  $w\textrm{-th}$ subkey is secure in (E4), it is also secure in the actual protocol. We refer to (E4) as the \textit{$w\textrm{-th}$ alternative scenario}.
	
	\subsection{Security of the $w\textrm{-th}$ subkey}
	
	{As {a consequence of the reduction above}, when proving the security of the $w\textrm{-th}$ subkey, we can assume that, in the $w\textrm{-group}$ rounds, Alice generates the characterised and uncorrelated states $\{\rho_{\rm model}^{\mu}\}_{\mu}$. Thanks to this, it becomes straightforward to prove its security using numerical methods. In particular, flexible techniques based on semidefinite programming (SDP) have been recently proposed {\cite{colesNumericalApproach2016,winickReliableNumerical2018,primaatmajaVersatileSecurity2019,bunandarNumericalFinitekey2020,georgeNumericalCalculations2021, zhouNumericalMethod2022,upadhyayaDimensionReduction2021,naharImperfectPhaseRandomisation2023a}}, which can handle almost any scenario, as long as the emitted states are characterised and uncorrelated, making them well suited to our purpose. The specific approach that we have developed uses ideas from these works but is targeted to this particular scenario. Below, {we provide an} overview of the main ideas, {and refer the reader to {\cref{app:SDP}} for a detailed description.}}
	
	{Each of Alice's generated states $\{\rho_{\rm model}^{\mu}\}_{\mu}$ can be diagonalised as}
	\begin{equation}
		\label{eq:rho_model_eigendec}
		\rho_{\rm model}^{\mu} = \sum_{n=0}^{\infty} p_{\lambda_n\vert \mu} \ketbra*{\lambda_n^{\mu}},
	\end{equation}
	{where we have omitted the dependence of the eigenvalues and eigenstates on $q$ for notational simplicity. Each set of eigenstates $\{\ket{\lambda_n^{\mu}}\}_n$ forms an orthonormal basis of the Fock space, and can be regarded as imperfect versions of the Fock states $\{\ket{n}\}_n$, with {the} two sets of states {converging} as $q\to 1$. Similarly, the eigenvalues $\{p_{\lambda_n\vert \mu}\}_n$ approach {a} Poisson distribution when $q\to 1$. Note that, when $q\neq 1$, the states $\{\ket{\lambda_n^{\mu}}\}_n$ depend slightly on the intensity setting $\mu$, and therefore the standard decoy-state method cannot be applied to this scenario. However, we can still} assume a counterfactual scenario in which Alice holds the ancillary system that purifies $\rho_{\rm model}^{\mu}$ and measures it to learn the value of $n$ for each round. {The information leakage of the $w\textrm{-th}$ sifted subkey can then be determined by estimating the fraction $q_{\lambda_1,w}$ of its bits that originated from emissions of $\ket{\lambda_1^{\mu_s}}$, and the phase-error rate $e^{\lambda_1,\mu_s}_{\textrm{ph},w}$ of these bits{, as shown in \cref{app:security}}. The first can be expressed as
	\begin{equation}
		\label{eq:q_t1}
		q_{\lambda_1,w} = \frac{p_{\lambda_1 \vert \mu_s} Y_{\lambda_1,\mu_s}^{Z,w}}{Q_{\mu_s,w}^{Z}},
	\end{equation}
	 where $Y_{\lambda_1,\mu_s}^{Z,w}$ is the yield probability of $\ket*{\lambda_1^{\mu_s}}$ when encoded in the $Z$ basis, which needs to be estimated, and $Q_{\mu_s,w}^{Z}$ is the observed rate at which Bob obtains detections conditioned on Alice choosing the intensity $\mu_s$, both users choosing the $Z$ basis, and the round being in $\mathcal{G}_w$. On the other hand,} to define the phase-error rate, we consider that, in the rounds in which both users choose the $Z$ basis and Alice prepares $\ket{\lambda_1^{\mu_s}}$, she {actually} generates the entangled state
	 \begin{equation}
	 	\label{eq:psi_lambda_t}
	 	\ket{\Psi_{Z}} = \frac{1}{\sqrt{2}} \left( \ket{0_Z}_A \hat{V}_{0_Z} \ket{\lambda_1^{\mu_s}} +  \ket{1_Z}_A \hat{V}_{1_Z}\ket{\lambda_1^{\mu_s}} \right),
	 \end{equation}
	 and performs an $X$-basis measurement on system $A$\;\cite{koashiSimpleSecurity2009}. Equivalently, she emits \begin{equation}
	 	\ket*{\lambda_{\textrm{vir}\beta}} \propto \ket*{\tilde\lambda_{\textrm{vir}\beta}} = \prescript{}{A}{\braket{\beta_X}{\Psi_{Z}}} = \frac{1}{2}( \hat{V}_{0_Z} + (-1)^{\beta}  \hat{V}_{1_Z})\ket{\lambda_1^{\mu_s}}
	 \end{equation} with probability $p_{\textrm{vir}\beta} = \norm*{\ket*{\tilde \lambda_{\textrm{vir}\beta}}}^2$, where $\beta \in \{0,1\}$ and $\ket{\beta_X} = (\ket{0_Z}+(-1)^{\beta} \ket{1_Z})/\sqrt{2}$. Also, we assume that Bob replaces his $Z$-basis measurement by an $X$-basis measurement, which is allowed due to {the basis-independent detection efficiency assumption}, (A8). The phase-error rate is then given by
	 \begin{equation}
	 	\label{eq:phase_error_rate}
	 	e^{\lambda_1,\mu_s}_{\textrm{ph},w} = \frac{p_{\textrm{vir}0} Y_{\textrm{vir}0}^{1_X} + p_{\textrm{vir}1} Y_{\textrm{vir}1}^{0_X}}{Y_{\lambda_1,\mu_s}^{Z,w}},
	 \end{equation}
	 where $Y_{\textrm{vir}\beta}^{(\beta \oplus 1)_X}$ is the probability that Bob obtains the measurement outcome $(\beta \oplus 1)_X$ conditioned on Alice emitting $\ket*{\lambda_{\textrm{vir}\beta}}$. 
	 
	 {In {\cref{app:SDP}}, we show how to obtain {a lower bound} $Y_{\lambda_1,\mu_s}^{Z,w,\textrm{L}}$ and {an upper bound} $e^{\lambda_1,\mu_s,\textrm{U}}_{\textrm{ph},w}$ on $Y_{\lambda_1,\mu_s}^{Z,w}$ and $e^{\lambda_1,\mu_s}_{\textrm{ph},w}$, respectively, using SDP techniques. In doing so, the main hurdle to overcome is the fact that the states $\{\rho_{\rm model}^{\mu}\}_{\mu}$ are infinite-dimensional, preventing us from finding their exact eigendecompositions using numerical methods, and from constructing finite-dimensional SDPs using these states. Instead, we construct the SDPs using the finite projections of $\{\rho_{\rm model}^{\mu}\}_{\mu}$} onto the subspace with up to $M$ photons\;{\cite{naharDecoyStateQuantum2022,naharImperfectPhaseRandomisation2023a}}, after numerically obtaining the eigendecompositions
	 \begin{equation}
	 	\Pi_M \rho_{\rm model}^{\mu} \Pi_M = \sum_{n=0}^{M} 
	 	p'_{\lambda_n\vert \mu} \ketbra*{\lambda_n^{\prime \mu}},
	 \end{equation}
	 {where $ \Pi_M = \sum_{n=0}^{M} \ketbra{n}$}. Then, by bounding the deviation between the eigenvalues and eigenvectors of $\rho_{\rm model}^{\mu}$ and $\Pi_M \rho_{\rm model}^{\mu} \Pi_M$ using perturbation theory results{, we can correct the SDP constraints and solutions, ensuring that the final bounds $Y_{\lambda_1,\mu_s}^{Z,w,\textrm{L}}$ and $e^{\lambda_1,\textrm{U}}_{\textrm{ph},w}$ apply to the original infinite-dimensional scenario. The secret-key rate obtainable per emitted $w\textrm{-group}$ pulse is then} given by
	 \begin{equation}
	 	\label{eq:skr}
	 	(p'_{\lambda_1\vert \mu_s} - \epsilon_{\rm val}^{\mu_s}) Y_{\lambda_1,\mu_s}^{Z,w,\textrm{L}} \big[1-h(e^{\lambda_1,\mu_s,\textrm{U}}_{\textrm{ph},w})\big] - Q_{\mu_s,w}^Z f h(E_{\mu_s,w}^{Z}),
	 \end{equation}
	 where $E_{\mu_s,w}^{Z}$ is the bit-error rate of the $w\textrm{-th}$ sifted subkey, $\epsilon_{\rm val}^{\mu_s} = 2 \sqrt{1-\textrm{Tr}[\Pi_M \rho_{\rm model}^{\mu_s} \Pi_M]}$ {is a correction term due to the finite projection}, $h(x)$ is the binary entropy function, $f$ is the error correction inefficiency{, and the rest of terms have already been defined.}

	 \section{Discussion}

 	We have proven the security of decoy-state QKD {in the presence of phase correlations}, which appear when {running} gain-switched laser sources at high-repetition rates. For simplicity, we have focused on the BB84 protocol, although our analysis can be straightforwardly adapted to other schemes, such as the three-state protocol\;\cite{boileauUnconditionalSecurity2005, tamakiLosstolerantQuantum2014} and measurement-device-independent QKD\;\cite{loMeasurementdeviceindependentQuantum2012}{, and our techniques may be applicable to other quantum communication protocols that rely on phase-randomised weak coherent sources, such as blind quantum computing \cite{dunjkoBlindQuantum2012} and quantum coin flipping \cite{pappaExperimentalPlug2014}.}  {Our proof requires knowledge of the parameters $l_c$ and $q$, see \cref{eq:assumption_lc,eq:assumption_q}.} {The former is an upper bound on the correlation length {(in a {generalised} Markovian sense)}, while the latter can be regarded as a lower bound on the uniformity of the conditional distribution of each phase given knowledge of all the other phases.  } 
 	
\begin{figure}[ht]
 		\centering
 		\includegraphics[width=0.6\columnwidth]{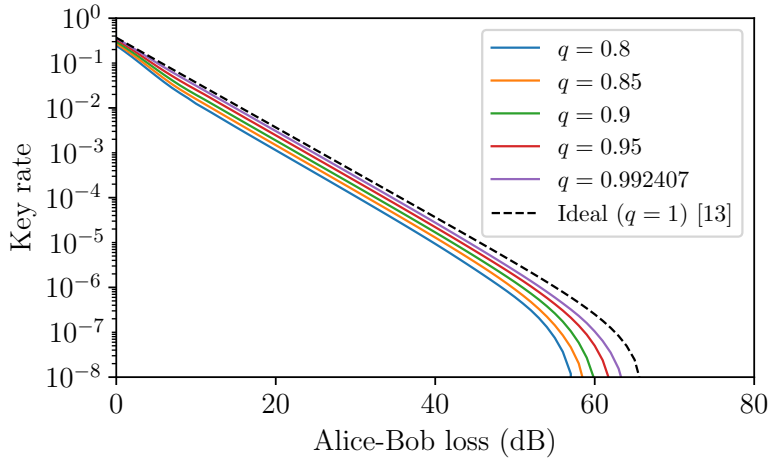}
 		\caption{Asymptotic secret-key rate of the decoy-state BB84 protocol with imperfect phase randomisation as a function of the overall system loss (solid lines), compared with the case of ideal phase randomisation\;\cite{maPracticalDecoy2005} (dashed line). We assume three intensities $\mu_s>\mu_w>\mu_v = 0$. Moreover, for simplicity, we set $\mu_w = \mu_s/5$, and optimise over $\mu_s$; while for the ideal case, we optimise over both $\mu_s$ and $\mu_w$. We consider a dark count probability $p_d = 10^{-8}$ for Bob's detectors, and an error correction inefficiency $f=1.16$.}
 		\label{fig:graph}
 	\end{figure}

 		 In \cref{fig:graph}, we plot the {overall} secret-key rate obtainable for {different} values of $q$. 
 	{We note that the {asymptotic} key rate does not depend on $l_c$, since it is only affected by the form of the states $\{\rho_{\rm model}^{\mu}\}_{\mu}$, which is independent of $l_c$; see \cref{eq:rho_model_def}.} To compute these results, we have used a simple channel model in which the only source of error is the dark count rate of Bob's detectors. Moreover, for simplicity, we have assumed that $\{\hat{V}_{b_\omega}\}$ are ideal BB84 encoding operators, and set $M = 9$.
 	
 	{To gauge the values of $q$ that one may expect in practical implementations, we examine the available literature.} {Recent} works \cite{kobayashiEvaluationPhase2014,grunenfelderPerformanceSecurity2020} {have studied the magnitude and properties of phase correlations in gain-switched lasers  under the implicit assumption that $l_c=1$}. In particular, Ref.\;\cite{kobayashiEvaluationPhase2014} argues that the phase difference between {adjacent} pulses follows a Gaussian distribution, {and shows how to} estimate its variance by measuring the fringe visibility $V$ in an asymmetric interferometer configuration. Under these assumptions, one can also calculate $q$ from the observed {visibility, see {\cref{app:experimental_q}}}. In particular, the value $V = 0.0019$ {recently} measured in Ref.\;\cite{grunenfelderPerformanceSecurity2020} for a {state-of-the-art} 5 GHz source corresponds to $q= 0.992407${; in \cref{fig:graph}, we have {included} the key rate obtainable for this value{, which is quite close to that of the ideal scenario.}}
 	
 	{While $l_c=1$ {might be} a good approximation to the phase distribution of {many} gain-switched laser {sources}, non-negligible correlations {could in principle} exist beyond immediately adjacent pulses{, especially in high-speed setups}. {Further work is needed to develop characterisation tests that can rigorously determine the value of $l_c$ and $q$ for any implementation.} {Since} the {asymptotic} key rate {offered by our proof} is {robust when decreasing} the value of $q$, {as evidenced by \cref{fig:graph},} {and independent of $l_c$,} 
 	{it is well placed to} guarantee the security of practical implementations while retaining {key rates close to the ideal scenario}, and we hope that the present {paper} will stimulate the experimental interest required to achieve this goal.}
 
 	\section*{Note}
 The security of decoy-state QKD with {imperfect phase randomisation} has also been recently investigated in Refs.\;\cite{naharQuantumKey2021,naharDecoyStateQuantum2022}. These works introduced insightful ideas that sparked the development of our security proof, and we recognise these important contributions. That being said, their security analysis contain{s} some conceptual flaws {that invalidate its application in the presence of phase correlations; see} {\cref{app:on_previous_results}}. {We note that the claims made in Refs.\;\cite{naharQuantumKey2021,naharDecoyStateQuantum2022} have been amended in \cite{naharImperfectPhaseRandomisation2023a}.}

	 \section*{Data availability statement 	 }

No new data were created or analysed in this study.

	 \section*{Statement}

This is the Accepted Manuscript version of an article accepted for publication in \textit{Quantum Science and Technology}. IOP Publishing Ltd is not responsible for any errors or omissions in this version of the manuscript or any version derived from it. This Accepted Manuscript is published under a CC BY licence. The Version of Record is available online at \href{https://doi.org/10.1088/2058-9565/ad141c}{10.1088/2058-9565/ad141c}.

\begin{acknowledgments}
	We thank Margarida Pereira and Víctor Zapatero for insightful discussions. G.C.-L.\ and M.C.\ acknowledge support by Cisco Systems Inc., the Galician Regional Government (consolidation of Research Units:\ AtlantTIC), the Spanish Ministry of Economy and Competitiveness (MINECO), the Fondo Europeo de Desarrollo Regional (FEDER) through Grant Number PID2020-118178RB-C21 and MICIN with funding from the European Union NextGenerationEU (PRTR-C17.I1), the Galician Regional Government with own funding through the "Planes Complementarios de I+D+I con las Comunidades Autónomas" in Quantum Communication, and the European Union's Horizon Europe research and innovation programme under the project QSNP (Quantum encryption and future quantum network technologies). G.C.-L.\ also acknowledges support from JSPS Postdoctoral Fellowships for Research in Japan. S.N.\ and N.L.\ acknowledge support from the Institute for Quantum Computing at the University of
	Waterloo through Innovation, Science, and Economic Development Canada, and the NSERC under the Discovery Grants Program, Grant No.\ 341495. K.T.\ acknowledges support from JSPS KAKENHI Grant Number JP18H05237 and JST-CREST JPMJCR 1671.
\end{acknowledgments}
 
 	\appendix
 	\renewcommand{\appendixname}{APPENDIX}
 	\section{Proof of \texorpdfstring{\cref{eq:cond_indep,eq:condition_w_group}}{Eqs. (4) and (7)}}
 	\label{app:proof_of_independence}

 	Although these are relatively straightforward consequences of Assumptions (A1) and (A2), for completeness, here we prove \cref{eq:cond_indep,eq:condition_w_group}.

 	Below we prove that, as a consequence of Assumption (A1), 
 	\begin{equation}
 		\label{eq:condition_objective}
 		f(\phi_i \vert  \phi_{N} ...  \phi_{i+1} \phi_{i-1} ... \phi_1) = f(\phi_i \vert \phi_{i+l_c} ...  \phi_{i+1}  \phi_{i-1} ... \phi_{i-l_c}),
 	\end{equation}
 	{which is the equality in \cref{eq:assumption_q}.}
 	Let $\mathcal{G}_w^{\neg i}$ be the set of rounds in $\mathcal{G}_{w}$, except the $i$-th round. We have that{, for all $i \in \mathcal{G}_{w}$,
 		\begin{equation}
 			\label{eq:proof_XYZ}
 			\begin{gathered}
 				f(\phi_i \vert  \vec \phi_{\mathcal{G}_{\overline w}}) = \int d\vec\phi_{\mathcal{G}_{w}^{\neg i}} f(\vec\phi_{\mathcal{G}_{w}^{\neg i}}) f(\phi_i \vert \vec \phi_{\mathcal{G}_{\overline w}\mathcal{G}_{w}^{\neg i}}) = \int d\vec\phi_{\mathcal{G}_{w}^{\neg i}} f(\vec\phi_{\mathcal{G}_{w}^{\neg i}}) f(\phi_i \vert\phi_{i+l_c} ...  \phi_{i+1}  \phi_{i-1} ... \phi_{i-l_c}) \\
 				= f(\phi_i \vert \phi_{i+l_c} ...  \phi_{i+1}  \phi_{i-1} ... \phi_{i-l_c}).
 			\end{gathered}
 		\end{equation}
 		where in the second to last equality we have used $\mathcal{G}_{\overline w}\mathcal{G}_{w}^{\neg i} =  \{{N}, ...,  {i+1}, {i-1}, ..., 1\}$ and \cref{eq:condition_objective}; and in the last equality we have used $\{ {i+l_c}, ...,  {i+1},  {i-1}, ..., {i-l_c}\} \notin \mathcal{G}_w^{\neg i}$.
 		Combining \cref{eq:condition_objective,eq:proof_XYZ}, we obtain
 		\begin{equation}
 			\begin{gathered}
 				f(\phi_i \vert \phi_{N} ... \phi_{i+1} \phi_{i-1}... \phi_{1}) 
 				= f(\phi_i|\vec{\phi}_{\mathcal{G}_{\overline w}}).
 			\end{gathered}
 		\end{equation}
 		This implies that the phases in $\mathcal{G}_w$ are conditionally independent of each other given knowledge of the phases in $\mathcal{G}_{\overline w}$, i.e.\ \cref{eq:cond_indep}. Also, combining \cref{eq:proof_XYZ} and Assumption~(A2), we obtain \cref{eq:condition_w_group}.
 		\subsection*{Proof of \texorpdfstring{\cref{eq:condition_objective}}{Eq. (A1)}}
 		\vspace{-20pt}
 		\begin{equation}
 			\begin{gathered}
 				f(\phi_i \vert  \phi_{N} ...  \phi_{i+1} \phi_{i-1} ... \phi_1)  = \frac{f(\phi_{N} ...  \phi_1)}{f(\phi_{N} ...  \phi_{i+1} \phi_{i-1} ... \phi_1)} \\
 				= \frac{f(\phi_N\vert \phi_{N-1} ... \phi_{1}) f(\phi_{N-1} \vert \phi_{N-2} ... \phi_{1})... f(\phi_{i+l_c + 1} \vert \phi_{i+l_c}... \phi_{1})  f(\phi_{i+l_c}...\phi_{1})}{f(\phi_N\vert \phi_{N-1} ... \phi_{i+1}\phi_{i-1}...\phi_{1}) f(\phi_{N-1} \vert \phi_{N-2} ... \phi_{i+1}\phi_{i-1}...\phi_{1})... f(\phi_{i+l_c + 1} \vert \phi_{i+l_c}...\phi_{i+1}\phi_{i-1}...\phi_{1}) f(\phi_{i+l_c}...\phi_{i+1}\phi_{i-1}...\phi_{1})} \\
 				\stackrel{(\times)}{=} \frac{f(\phi_N\vert \phi_{N-1} ... \phi_{N-l_c}) f(\phi_{N-1} \vert \phi_{N-2} ... \phi_{N-l_c-1})... f(\phi_{i+l_c + 1} \vert \phi_{i+l_c}... \phi_{i+1}) f(\phi_{i+l_c}...\phi_{1})}{f(\phi_N\vert \phi_{N-1} ... \phi_{N-l_c}) f(\phi_{N-1} \vert \phi_{N-2} ... \phi_{N-l_c-1})... f(\phi_{i+l_c + 1} \vert \phi_{i+l_c}... \phi_{i+1}) f(\phi_{i+l_c}...\phi_{i+1}\phi_{i-1}...\phi_{1})} \\
 				= \frac{f(\phi_{i+l_c}...\phi_{1})}{f(\phi_{i+l_c}...\phi_{i+1}\phi_{i-1}...\phi_{1})} \\
 				= \frac{f(\phi_1 \vert \phi_2 ... \phi_{i+l_c})f(\phi_2 \vert \phi_3 ... \phi_{i+l_c})... f(\phi_{i-l_c-1} \vert \phi_{i-l_c}...\phi_{i+l_c}) f(\phi_{i-l_c}...\phi_{i+l_c})}{f(\phi_1 \vert \phi_2 ...\phi_{i-1}\phi_{i+1}... \phi_{i+l_c})f(\phi_2 \vert \phi_3...\phi_{i-1}\phi_{i+1} ... \phi_{i+l_c})... f(\phi_{i-l_c-1} \vert \phi_{i-l_c}...\phi_{i-1}\phi_{i+1}...\phi_{i+l_c}) f(\phi_{i-l_c}...\phi_{i-1}\phi_{i+1}...\phi_{i+l_c})}\\
 				\stackrel{(*)}{=} \frac{f(\phi_1 \vert \phi_2 ... \phi_{1+l_c})f(\phi_2 \vert \phi_3 ... \phi_{2+l_c})... f(\phi_{i-l_c-1} \vert \phi_{i-l_c}...\phi_{i}) f(\phi_{i-l_c}...\phi_{i+l_c})}{f(\phi_1 \vert \phi_2 ... \phi_{1+l_c})f(\phi_2 \vert \phi_3...\phi_{2+l_c})... f(\phi_{i-l_c-1} \vert \phi_{i-l_c}...\phi_{i}) f(\phi_{i-l_c}...\phi_{i-1}\phi_{i+1}...\phi_{i+l_c})}\\
 				= \frac{f(\phi_{i-l_c}...\phi_{i+l_c})}{f(\phi_{i-l_c}...\phi_{i-1}\phi_{i+1}...\phi_{i+l_c})} \\
 				= f(\phi_i \vert \phi_{i+l_c} ...  \phi_{i+1}  \phi_{i-1} ... \phi_{i-l_c}),
 			\end{gathered}
 		\end{equation}
 		where in the equality marked by an asterisk, we have used
 		\begin{equation}
 			\begin{gathered}
 				f(\phi_j \vert \phi_{j+1}... \phi_{J}) = \frac{f(\phi_j... \phi_{J})}{f(\phi_{j+1}... \phi_{J})} \\
 				= \frac{f(\phi_{J} \vert \phi_{J-1}...\phi_{j})f(\phi_{J-1} \vert \phi_{J-2}...\phi_{j})...f(\phi_{j+l_c+1} \vert \phi_{j+l_c}...\phi_j) f(\phi_{j+l_c}...\phi_j)}{f(\phi_{J} \vert \phi_{J-1}...\phi_{j+1})f(\phi_{J-1} \vert \phi_{J-2}...\phi_{j+1})...f(\phi_{j+l_c+1} \vert \phi_{j+l_c}...\phi_{j+1}) f(\phi_{j+l_c}...\phi_{j+1})} \\
 				\stackrel{(\times)}{=}  \frac{f(\phi_{J} \vert \phi_{J-1}...\phi_{J-l_c})f(\phi_{J-1} \vert \phi_{J-2}...\phi_{J-l_c-1})...f(\phi_{j+l_c+1} \vert \phi_{j+l_c}...\phi_{j+1}) f(\phi_{j+l_c}...\phi_j)}{f(\phi_{J} \vert \phi_{J-1}...\phi_{J-l_c})f(\phi_{J-1} \vert \phi_{J-2}...\phi_{J-l_c-1})...f(\phi_{j+l_c+1} \vert \phi_{j+l_c}...\phi_{j+1}) f(\phi_{j+l_c}...\phi_{j+1})} \\
 				= \frac{f(\phi_{j+l_c}...\phi_j)}{f(\phi_{j+l_c}...\phi_{j+1})} = f(\phi_j \vert \phi_{j+1}... \phi_{j+l_c}),
 			\end{gathered}
 		\end{equation}
 		and in the equalities marked by a cross, we have used Assumption (A1).

 	\section{Obtaining the required bounds using SDPs}
 	\label{app:SDP}
 	
 	Here, we show how to obtain the bounds 
 	$q_{\lambda_1,w}^{\rm L}$ and $e^{\lambda_1,\mu_s,\textrm{U}}_{\textrm{ph},w}$ using semidefinite programming techniques, and employ these to derive an asymptotic lower bound on the secret-key rate. To do so, for simplicity, we assume that Eve performs a collective attack. However, the set of bounds we obtain, and thus the overall security proof, is also valid for general attacks, due to the extension of the quantum de Finetti theorem \cite{rennerSymmetryLarge2007}
 	to infinite-dimensional systems \cite{rennerFinettiRepresentation2009}. {We note that, as an alternative to the SDP approach presented here, {which uses ideas from Refs.~\cite{naharDecoyStateQuantum2022,naharImperfectPhaseRandomisation2023a},} one could also obtain these bounds using linear programming techniques, by using the trace distance inequality to account for the dependence of the eigenstates $\ket{\lambda_n^{\mu}}$ on the intensity $\mu$ (see Refs.\;\cite{caoDiscretephaserandomizedCoherent2015,tamakiDecoystateQuantum2016,wangFinitekeySecurity2018,wangMeasurementdeviceindependentQuantum2021}). However, according to our preliminary numerical simulations, this would result in much more pessimistic bounds.} 
 	
 	Eve's collective attack can be described as a quantum channel $\Lambda$ acting separately on each of Alice's emitted photonic systems. Let us assume that, in a given round, Bob performs a POVM that contains some element $\Gamma$. The probability that Bob obtains the outcome associated to $\Gamma$ when Alice sends him a quantum state $\sigma$ can be expressed as
 	\begin{equation}
 		\label{eq:reduction_1}
 		\Tr[\Lambda(\sigma) \Gamma ] = \Tr[\sum_{l}  E_l \sigma  E^{\dagger}_l \Gamma] =  \sum_{l} \Tr[ E_l \sigma  E^{\dagger}_l \Gamma ] = \sum_{l} \Tr[\sigma  E^{\dagger}_l \Gamma E_l ] = \Tr[\sigma \sum_{l}  E^{\dagger}_l \Gamma E_l ] = \Tr[\sigma  H ],
 	\end{equation}
 	where $\{E_l\}$ are the set of Kraus operators of the operator-sum representation \cite{nielsenQuantumComputation2011} for the channel $\Lambda$, and
 	\begin{equation}
 		0\leq  H \coloneqq \sum_{l}  E^{\dagger}_l \Gamma E_l  \leq \sum_{l}  E^{\dagger}_l  E_l = \mathbb{I}.
 	\end{equation}
 	
 	We denote Bob's $Z$ and $X$ basis POVMs as, respectively, $\{\Gamma_{0_Z}, \Gamma_{1_Z}, \Gamma_{f}\}$ and $\{\Gamma_{0_X}, \Gamma_{1_X}, \Gamma_{f}\}$. Note that the element associated to an inconclusive result, $\Gamma_{f}$, is the same for both bases, due to Assumption (A8) (basis-independent detection efficiency).

 	\subsection{Lower bound on \texorpdfstring{$q_{\lambda_1,w}$}{qlambda1w}}
 	
 	To estimate the fraction $q_{\lambda_1,w}$, we need to estimate the yield $Y_{\lambda_1,\mu_s}^{Z,w}${, see \cref{eq:q_t1}.} Substituting $\sigma \to \Big(\tfrac{1}{2} \hat{V}_{0_Z} \rho \hat{V}_{0_Z}^{\dagger} + \tfrac{1}{2} \hat{V}_{1_Z} \rho \hat{V}_{1_Z}^{\dagger}\Big)$ and $\Gamma \to (\Gamma_{0_Z} + \Gamma_{1_Z})$ in \cref{eq:reduction_1}, we obtain
 	\begin{equation}
 		\label{eq:reduction_2}
 		\begin{gathered}
 			\Tr[\Lambda\Big(\tfrac{1}{2}  \hat{V}_{0_Z} \rho  \hat{V}_{0_Z}^{\dagger} + \tfrac{1}{2}  \hat{V}_{1_Z} \rho  \hat{V}_{1_Z}^{\dagger}\Big)(\Gamma_{0_Z}+\Gamma_{1_Z})] = \Tr[\left(\tfrac{1}{2}  \hat{V}_{0_Z} \rho  \hat{V}_{0_Z}^{\dagger} + \tfrac{1}{2}  \hat{V}_{1_Z} \rho  \hat{V}_{1_Z}^{\dagger}\right) {H} ]\\ = \Tr[\tfrac{1}{2}\rho  \hat{V}_{0_Z}^{\dagger} {H}  \hat{V}_{0_Z} + \tfrac{1}{2}\rho  \hat{V}_{1_Z}^{\dagger} {H}  \hat{V}_{1_Z} ]  = \Tr[\rho \tfrac{1}{2} (  \hat{V}_{0_Z}^{\dagger} {H}  \hat{V}_{0_Z} +  \hat{V}_{1_Z}^{\dagger} {H}  \hat{V}_{1_Z})] = \Tr[\rho  J].
 		\end{gathered}
 	\end{equation}
 	where we have defined
 	\begin{equation}
 		0 \leq  J \coloneqq \tfrac{1}{2} (  \hat{V}_{0_Z}^{\dagger} {H}  \hat{V}_{0_Z} +  \hat{V}_{1_Z}^{\dagger} {H}  \hat{V}_{1_Z}) \leq \tfrac{1}{2} (  \hat{V}_{0_Z}^{\dagger} \hat{V}_{0_Z} +  \hat{V}_{1_Z}^{\dagger}  \hat{V}_{1_Z}) =\mathbb{I}.
 	\end{equation}
 	Substituting first $\rho \to \ketbra{\lambda_1^{\mu_s}}$ and then $\rho \to \rho_{\rm model}^{\mu}$ in \cref{eq:reduction_2}, we obtain
 	\begin{equation}
 		\begin{gathered}
 			Y_{\lambda_1,\mu_s}^{Z,w} = \Tr[\ketbra{\lambda_1^{\mu_s}} J] \\ 
 			Q_{\mu,w}^{Z} = \Tr[\rho_{\rm model}^{\mu} J].
 		\end{gathered}
 	\end{equation}
 	This implies that we can express a lower bound on $Y_{\lambda_1,\mu_s}^{Z,w}$ {as the SDP
 		\begin{equation}
 			\begin{alignedat}{3}
 				\label{eq:fraction_sdp}
 				&\!\!\min_{J}\hspace{0.4em}&&\Tr[\ketbra{ \lambda_1^{\mu_s}} J] \\
 				& \!\!\;\textrm{s.t.} &&\Tr[\rho_{\rm model}^{\mu} J] = Q_{\mu,w}^{Z}, \quad &&\forall \mu \\
 				& &&\;0 \leq J \leq \mathbb{I}.
 			\end{alignedat}
 		\end{equation}
 		However, as explained in the main text}, one cannot solve this SDP numerically because (1) it is infinitely dimensional and (2) the eigendecomposition of $\rho_{\rm model}^{\mu}$ is {unknown}. To {overcome these problems}, we consider the projection of {the} state {$\rho_{\rm model}^{\mu}$} onto the subspace with up to $M$ photons, and numerically find its eigendecomposition,
 	\begin{equation}
 		\label{eq:rho_model_prime_def}
 		\rho_{\rm model}^{\prime \mu} = \frac{\Pi_M \rho_{\rm model}^{\mu} \Pi_M }{\Tr[\Pi_M \rho_{\rm model}^{\mu} \Pi_M]} = \sum_{n=0}^{M} 
 		\frac{p'_{\lambda_n\vert \mu}}{{\Tr[\Pi_M \rho_{\rm model}^{\mu} \Pi_M]}} \ketbra*{\lambda_n^{\prime \mu}},
 	\end{equation}
 	%
 	where the decomposition has $M+1$ terms because the projection is in a space of dimension $M+1$. 
 	
 	{The objective is to construct a relaxed version of \cref{eq:fraction_sdp} using the finite-dimensional states $\rho_{\rm model}^{\prime \mu}$ and $\ket*{\lambda_1^{\prime \mu_s}}$ rather than their infinite-dimensional counterparts. To do so,} we make use of the following results {from \cite{naharDecoyStateQuantum2022}}
 	\begin{gather}
 		F(\rho_{\rm model}^{\mu},\rho_{\rm model}^{\prime \mu}) = F_{\rm proj}^{\mu}, \label{eq:proj_bound} \\
 		\abs*{p_{\lambda_n\vert \mu} - p'_{\lambda_n\vert \mu}} \leq \epsilon_{\rm val}^{\mu},  \label{eq:eigenval_bound} \\ 
 		\abs*{\braket*{\lambda_n^{\prime \mu}}{\lambda_n^{\mu}}}^2  \geq F_{\textrm{vec},\lambda_n}^{\mu}, \label{eq:eigenvec_bound}
 	\end{gather}
 	where $F(\sigma,\sigma')$ is the fidelity between $\sigma$ and $\sigma'${, given by}
 	\begin{equation}
 		F(\sigma,\sigma') = \Tr[\sqrt{\sqrt{\sigma}\sigma'\sqrt{\sigma}}]^2,
 	\end{equation}
 	and  $F_{\rm proj}^{\mu},\epsilon_{\rm val}^{\mu},F_{\textrm{vec},\lambda_n}^{\mu} \in [0,1]$ {are given by
 	\begin{gather}
 		F_{\rm proj}^{\mu} \coloneqq  \sum_{n=0}^{M}  p'_{\lambda_n\vert \mu}, \\ 
 		 	\epsilon_{\rm val}^{\mu} \eqqcolon 2 \sqrt{1-F_{\rm proj}^{\mu}}, \\
 		 		\abs*{\braket*{\lambda_n^{\prime \mu}}{\lambda_n^{\mu}}}^2  \geq 1-\left(\frac{\epsilon_{\rm val}^{\mu}}{\delta_n}\right)^2 \coloneqq F_{\textrm{vec},\lambda_n}^{\mu}. \label{eq:F_vec}
 	\end{gather}}
 		 	In \cref{eq:F_vec}, $\delta_0 = p'_{\lambda_{0}\vert \mu} -  p'_{\lambda_{1}\vert \mu} - \epsilon_{\rm val}^{\mu}$ and for $n>1$,
 		 	\begin{equation}
 		 		\delta_n = \min \{p'_{\lambda_{n-1}\vert \mu} - p'_{\lambda_{n}\vert \mu} - \epsilon_{\rm val}^{\mu},\, p'_{\lambda_{n}\vert \mu} -  p'_{\lambda_{n+1}\vert \mu} - \epsilon_{\rm val}^{\mu}\},
 		 	\end{equation}
 	Also, we use the following inequality
 	\begin{equation}
 		\label{eq:G_function}
 		G_{-}(\textrm{Tr}[\sigma'  M], F(\sigma,\sigma')) \leq  \textrm{Tr}[\sigma  M] \leq G_{+}(\textrm{Tr}[\sigma'  M],F(\sigma,\sigma')),
 	\end{equation}
 	which holds for any two density {operators} $\sigma,\sigma'$ and any $0\leq  M \leq \mathbb{I}$, and where
 	\begin{equation}
 		G_-(y,z) =
 		\begin{cases}
 			g_-(y,z)  & \quad \text{if } y > 1 -z \\
 			0  & \quad \text{otherwise}
 		\end{cases}
 		\quad \quad
 		\textrm{and}
 		\quad \quad
 		G_+(y,z) =
 		\begin{cases}
 			g_+(y,z)  & \quad \text{if } y < z \\
 			1  & \quad \text{otherwise}
 		\end{cases}
 	\end{equation}
 	with
 	\begin{equation}
 		g_{\pm} (y,z) = y + (1-z)(1-2y) \pm 2\sqrt{z(1-z)y(1-y)}.
 	\end{equation}
  The proofs for {the} results {in \cref{eq:proj_bound,eq:eigenval_bound,eq:eigenvec_bound,eq:G_function}} are {given in {\cref{appsec:bounds_section} below}}.
 	
 	Let $J^{\ast}$ be the operator that minimises the SDP in {\cref{eq:fraction_sdp}}. We have that
 	\begin{equation}
 		\label{eq:fraction_first_step}
 		Y_{\lambda_1,\mu_s}^{Z,w} \geq \ev*{J^{\ast}}{\lambda_1^{\mu_s}} \geq  G_{-}\big(\ev*{J^{\ast}}{\lambda_1^{\prime\mu_s}}, F_{\textrm{vec},\lambda_1}^{\mu_s}\big),
 	\end{equation}
 	where in the last inequality we have used \cref{eq:eigenvec_bound,eq:G_function} and the fact that $G_{-}$ is increasing with respect to its second argument. On the other hand, we have that
 	\begin{equation}
 		\label{eq:fraction_sdp_relationship}
 		\ev*{J^{\ast}}{\lambda_1^{\prime\mu_s}} \geq \ev*{J^{\ast \ast}}{\lambda_1^{\prime\mu_s}} \eqqcolon 	Y_{\lambda_1,\mu_s}^{\prime Z,w,\textrm{L}},
 	\end{equation}
 	where $Y_{\lambda_1,\mu_s}^{\prime Z,w,\textrm{L}}$ is the solution of the SDP
 	\begin{equation}
 		\begin{alignedat}{3}
 			\label{eq:fraction_sdp_finite}
 			&\!\!\min_{J}\hspace{0.4em}&&\Tr[\ketbra*{ \lambda_1^{\prime \mu_s}} J] \\
 			& \!\!\;\textrm{s.t.} &&G_{-} (Q_{\mu,w}^{Z},F_{\rm proj}^{\mu}) \leq \Tr[\rho_{\rm model}^{\prime \mu}J] \leq G_{+} (Q_{\mu,w}^{Z},F_{\rm proj}^{\mu}), \, \quad &&\forall \mu \\
 			& &&\;0 \leq J \leq \mathbb{I};
 		\end{alignedat}
 	\end{equation}
 	and {$J^{* *}$} is the operator that minimises this SDP. In \cref{eq:fraction_sdp_finite}{, $\rho_{\rm model}^{\prime \mu}$ {is given by} \cref{eq:rho_model_prime_def}, and in} the first inequality of \cref{eq:fraction_sdp_finite}, we have used \cref{eq:proj_bound,eq:G_function}. \Cref{eq:fraction_sdp_relationship} holds because the constraints of \cref{eq:fraction_sdp_finite} are looser than those of {\cref{eq:fraction_sdp}}, i.e.\ all operators that satisfy the constraints of {\cref{eq:fraction_sdp}}, including $J^{\ast}$, also satisfy the constraints of \cref{eq:fraction_sdp_finite}. Note that the {states $\rho_{\rm model}^{\prime \mu}$ and $\ket*{ \lambda_1^{\prime \mu_s}}$ live in the finite subspace spanned by $\{\ket{0},...,\ket{M}\}$, and therefore, the action of $J$ outside this finite subspace is irrelevant as far as the optimisation problem in \cref{eq:fraction_sdp_finite} is concerned. As a consequence,} we can restrict the optimisation search to operators $J$ that act only on this finite subspace, i.e.\ {\cref{eq:fraction_sdp_finite} is actually a finite-dimensional SDP that} we can solve numerically.
 	
 	Combining \cref{eq:fraction_first_step,eq:fraction_sdp_relationship}, and using the fact that $G_{-}$ is increasing with respect to its first argument, we obtain the bound
 	\begin{equation}
 		\label{eq:fraction_yield_bound}
 		Y_{\lambda_1,\mu_s}^{Z,w} \geq G_{-}\big(Y_{\lambda_1,\mu_s}^{\prime Z,w,\textrm{L}}, F_{\textrm{vec},\lambda_1}^{\mu_s}\big) \eqqcolon Y_{\lambda_1,\mu_s}^{Z,w,\textrm{L}}.
 	\end{equation}
 	Using \cref{eq:eigenval_bound,eq:fraction_yield_bound}, we finally obtain {the} bound 
 	\begin{equation}
 		\label{eq:q_1_L}
 		q_{\lambda_1,w} \geq \frac{(p'_{\lambda_n\vert \mu} - \epsilon_{\rm val}^{\mu_s}) Y_{\lambda_1,\mu_s}^{Z,w,\textrm{L}}}{Q_{\mu_s,w}^{Z}} \eqqcolon q_{\lambda_1,w}^{\rm L}.
 	\end{equation}
 	
 	\subsection{Upper bound on \texorpdfstring{$e^{\lambda_1,\mu_s}_{\textrm{ph},w}$}{elambda1phw}}
 	
 	
	The phase-error rate is given by {\cref{eq:phase_error_rate}.} We can express each term in the numerator of {this equation} as 
 	\begin{equation}
 		\begin{gathered}
 			p_{\textrm{vir}\beta} Y_{\textrm{vir}\beta}^{(\beta \oplus 1)_X} = p_{\textrm{vir}\beta}  \Tr[\Lambda\left(\ketbra*{\lambda_{\textrm{vir}\beta}}\right) \Gamma_{(\beta \oplus 1)_X}] \\= \Tr[ p_{\textrm{vir}\beta}  \ketbra*{\lambda_{\textrm{vir}\beta}}L_{(\beta \oplus 1)_X}] =  \Tr\big[\ketbra*{\tilde \lambda_{\textrm{vir}\beta}}L_{(\beta \oplus 1)_X}\big],
 		\end{gathered}
 	\end{equation}
 	where in the second equality we have {used \cref{eq:reduction_1} with the substitutions $\Gamma \to \Gamma_{(\beta \oplus 1)_X}$, $H \to L_{(\beta \oplus 1)_X}$ and $\sigma \to \ketbra*{\lambda_{\textrm{vir}\beta}}$}. By substituting $\sigma \to \hat{V}_{b_{\omega_A}} \rho_{\rm model}^{\mu} \hat{V}_{b_{\omega_A}}^{\dagger}$ instead, we obtain
 	\begin{equation}
 		Q_{\mu, b_{\omega_A},w}^{(\beta \oplus 1)_X} = \Tr\big[\hat{V}_{b_{\omega_A}} \rho_{\rm model}^{\mu} \hat{V}_{b_{\omega_A}}^{\dagger} L_{(\beta \oplus 1)_X}\big],
 	\end{equation}
 	{where $Q_{\mu, b_{\omega_A},w}^{(\beta \oplus 1)_X}$ is the observed rate at which Bob obtains the result $(\beta \oplus 1)_X$ conditioned on Alice choosing intensity $\mu$, basis $\omega_A$ and bit $b$, Bob choosing the $X$ basis, and the round being in $\mathcal{G}_w$.}
 	{This means that} an upper bound on $p_{\textrm{vir}\beta} Y_{\textrm{vir}\beta}^{(\beta \oplus 1)_X}$ can be expressed as the SDP 
 		\begin{equation}
 			\label{eq:ph_error_sdp}
 			\begin{alignedat}{3}
 				&\!\!\max_{L_{(\beta \oplus 1)_X}} \!&&\Tr\big[\ketbra*{\tilde\lambda_{\textrm{vir}\beta}} L_{(\beta \oplus 1)_X}\big] \\
 				&\,\,\,\textrm{s.t.} &&\Tr\big[ \hat{V}_{b_{\omega_A}} \rho_{\rm model}^{\mu}  \hat{V}_{b_{\omega_A}}^{\dagger}\! L_{(\beta \oplus 1)_X}\big] = Q_{\mu, b_{\omega_A},w}^{(\beta \oplus 1)_X}, \,\forall \mu,b,\omega_A \\
 				& &&\;0 \leq L_{(\beta \oplus 1)_X} \leq \mathbb{I}.
 			\end{alignedat}
 	\end{equation}
 	As before, we need to {find a finite-dimensional relaxation of \cref{eq:ph_error_sdp}}
  that we can solve numerically. Let $L_{(\beta \oplus 1)_X}^{\star}$ be the operator that maximises the SDP in {\cref{eq:ph_error_sdp}}, and let
 	\begin{equation}
 		{M}_{\rm ph} \coloneqq \ketbra{0_X} \otimes L_{1_X}^{\star} + \ketbra{1_X} \otimes L_{0_X}^{\star}.
 	\end{equation}
 	We have that
 	\begin{equation}
 		\label{eq:num_in_terms_of_Mph}
 		\begin{gathered}
 			p_{\textrm{vir}0} Y_{\textrm{vir}0}^{1_X} + p_{\textrm{vir}1}  Y_{\textrm{vir}1}^{0_X} \leq  \ev*{ L_{1_X}^{\star}}{\tilde \lambda_{\textrm{vir}0}} +  \ev*{ L_{0_X}^{\star}}{\tilde\lambda_{\textrm{vir}1}} 	= 	\ev*{{M}_{\rm ph}}{\Psi_{Z}},  
 		\end{gathered}
 	\end{equation}
 	{where $\ket*{{\Psi_{Z}}}$ is defined in \cref{eq:psi_lambda_t}}. Now, let us define the entangled state
 	\begin{equation}
 		\ket*{\Psi'_{Z}} = \frac{1}{\sqrt{2}} \left( \ket{0_Z}  \hat{V}_{0_Z} \ket{\lambda_1^{\prime \mu_s}} +  \ket{1_Z}  \hat{V}_{1_Z}\ket{\lambda_1^{\prime \mu_s}} \right).
 	\end{equation}
 	and the unnormalised states 
 	\begin{equation}
 		\label{eq:virtual_state_finite}
 		\ket*{\tilde\lambda'_{\textrm{vir}\beta}} = \braket*{\beta_X}{\Psi'_{Z}} = \frac{1}{2}( \hat{V}_{0_Z} + (-1)^{\beta}  \hat{V}_{1_Z})\ket*{\lambda_1^{\prime \mu_s}}.
 	\end{equation}
 	
 	We have that
 	\begin{equation}
 		\abs{\braket*{\Psi'_{Z}}{\Psi_{Z}}}^2 = \abs{\tfrac{1}{2} \mel{\lambda_1^{\prime \mu_s}}{\hat{V}_{0Z}^{\dagger} \hat{V}_{0Z}}{\lambda_1^{\mu_s}} + \tfrac{1}{2} \mel{\lambda_1^{\prime \mu_s}}{\hat{V}_{1Z}^{\dagger} \hat{V}_{1Z}}{\lambda_1^{\mu_s}}}^2 = \abs{\braket*{\lambda_1^{\prime \mu_s}}{\lambda_1^{\mu_s}}}^2 \geq F_{\textrm{vec},\lambda_1}^{\mu_s},
 	\end{equation}
 	where the inequality is due to \cref{eq:eigenvec_bound}. Therefore, applying the bound in \cref{eq:G_function}, and using the fact that $G_{+}$ is a decreasing function with respect to its second argument, 
 	\begin{equation}
 		\label{eq:step_2}
 		\ev*{{M}_{\rm ph}}{\Psi_{Z}} \leq G_{+}\big(\ev*{{M}_{\rm ph}}{\Psi'_{Z}}, F_{\textrm{vec},\lambda_1}^{\mu_s}\big).
 	\end{equation}
 	
 	On the other hand, we have that
 	\begin{equation}
 		\label{eq:step_3}
 		\begin{gathered}
 			\ev*{{M}_{\rm ph}}{\Psi'_{Z}} = \ev*{L_{1_X}^{\star}}{\tilde\lambda'_{\textrm{vir}0}} +  \ev*{ L_{0_X}^{\star}}{\tilde\lambda'_{\textrm{vir}1}} \\ \leq  \ev*{L_{1_X}^{\ast \ast}}{\tilde\lambda'_{\textrm{vir}0}} +  \ev*{ L_{0_X}^{\ast \ast}}{\tilde\lambda'_{\textrm{vir}1}} \eqqcolon \tilde Y_{\textrm{vir}0}^{\prime 1_X} + \tilde Y_{\textrm{vir}1}^{\prime 0_X},
 		\end{gathered}
 	\end{equation}
 	where $\tilde Y_{\textrm{vir}\beta}^{\prime (\beta \oplus 1)_X}$ is the solution {to} the {following} SDP
 	\begin{equation}
 		\label{eq:ph_error_sdp_finite}
 		\begin{alignedat}{3}
 			&\!\!\max_{L_{(\beta \oplus 1)_X}} \hspace{0.4em}&&\Tr\big[\ketbra*{\tilde \lambda'_{\textrm{vir}\beta}} L_{(\beta \oplus 1)_X}\big] \\
 			&\!\!\;\textrm{s.t.} &&G_{-} ( Q_{\mu, b_{\omega_A},w}^{(\beta \oplus 1)_X} ,F_{\rm proj}^{\mu}) \leq \Tr\big[\hat{V}_{b_{\omega_A}} \rho_{\rm model}^{\prime \mu} \hat{V}_{b_{\omega_A}}^{\dagger} L_{(\beta \oplus 1)_X}\big] \leq G_{+} ( Q_{\mu, b_{\omega_A},w}^{(\beta \oplus 1)_X} ,F_{\rm proj}^{\mu}), \quad &&\forall \mu,\omega_A,b \\
 			& &&\;0 \leq L_{(\beta \oplus 1)_X} \leq \mathbb{I};
 		\end{alignedat}
 	\end{equation}
 	and $L_{(\beta \oplus 1)_X}^{\ast \ast}$ is the operator that maximises this SDP. In \cref{eq:ph_error_sdp_finite}{, $\rho_{\rm model}^{\prime \mu}$ {is given by} \cref{eq:rho_model_prime_def}, and in the} first inequality of \cref{eq:ph_error_sdp_finite}, we have used \cref{eq:proj_bound,eq:G_function}. Note that the inequality in \cref{eq:step_3} holds because $L_{(\beta \oplus 1)_X}^{\ast}$ satisfies the constraints of \cref{eq:ph_error_sdp_finite}.
 		
 	Combining \cref{eq:num_in_terms_of_Mph,eq:step_2,eq:step_3}, and using the fact that $G_{+}$ is increasing with respect to its first argument, we obtain the bound
 	\begin{equation}
 		\label{eq:YtildephU}
 		p_{\textrm{vir}0} Y_{\textrm{vir}0}^{1_X} + p_{\textrm{vir}1}  Y_{\textrm{vir}1}^{0_X} \leq  G_{+}\big(\tilde Y_{\textrm{vir}0}^{\prime 1_X} + \tilde Y_{\textrm{vir}1}^{\prime 0_X}, F_{\textrm{vec},\lambda_1}^{\mu_s}\big) \eqqcolon \tilde Y_{\rm ph}^{\rm U}.
 	\end{equation}
 	Then, using \cref{eq:fraction_yield_bound,eq:YtildephU}, we finally obtain {the} bound on {the phase-error rate of the $w$-th sifted subkey},
 	\begin{equation}
 		\label{eq:eph_bound}
 		e^{\lambda_1,\mu_s}_{\textrm{ph},w} \leq \frac{\tilde Y_{\rm ph}^{\rm U}}{Y_{\lambda_1,\mu_s}^{Z,w,\textrm{L}}} \eqqcolon e^{\lambda_1,\mu_s,\textrm{U}}_{\textrm{ph},w}.
 	\end{equation}
 	
 	\subsection{Secret-key rate}
 	
 	{Putting all together}, a lower bound on the fraction of the $w$-th sifted {subkey} that can be turned into a secret key is given by
 	\begin{equation}
 		\label{eq:fraction_key_rate_w}
 		F_w \geq q_{\lambda_1,w}^{\rm L} \big[1-h(e_{{\rm ph},w}^{\lambda_1,\mu_s,\textrm{U}})\big] - f h(E^Z_{\mu_s, w})  \coloneqq F_w^{\rm L},
 	\end{equation}
 	where $E^Z_{\mu_s, w}$ is the error rate conditioned on Alice choosing the intensity $\mu_s$, both users choosing the $Z$ basis, and the round being in $\mathcal{G}_w$; and a lower bound on the secret-key rate obtainable per emitted $w$-group pulse is given by
 	\begin{equation}
 		R_w \geq p_{\mu_s} p_{Z_A} p_{Z_B} Q_{\mu_s,w}^{Z} F_w^{\rm L} \coloneqq R_w^{\rm L}.
 	\end{equation}
 	By assuming that $p_{\mu_s}$, $p_{Z_A}$ and $p_{Z_B}$ all approach one, which is optimal when $N \to \infty$, and substituting $q_{\lambda_1,w}^{\rm L}$ by its definition in \cref{eq:q_1_L}, we obtain \cref{eq:skr}.
 	
 {For completeness, we note that the procedure presented above can be used to obtain bounds on $q_{\lambda_n,w}$ and $e_{{\rm ph},w}^{\lambda_n,\mu_s}$ for any $n$, not just $n=1$. In fact, a more general lower bound on the fraction of the $w$-th sifted key that can be turned into a secret key is given by
 	 \begin{equation}
 		\label{eq:fraction_key_rate_w_2}
 		F_w \geq \sum_{n\in \mathcal{N}} q_{\lambda_n,w}^{\rm L} \big[1-h(e_{{\rm ph},w}^{\lambda_n,\mu_s,\textrm{U}})\big] - f h(E^Z_{\mu_s, w})  \coloneqq F_w^{\rm L},
 	\end{equation}
	where $\mathcal{N}$ denotes the set of values of $n$ for which one obtained bounds on $q_{\lambda_n,w}$ and $e_{{\rm ph},w}^{\lambda_n,\mu_s}$. According to our simulations, by obtaining bounds for $n=0$, one can obtain a small key-rate improvement in some scenarios (particularly, for low attenuations and relative low values of $q$), but we have not found any scenario in which one can obtain a positive key-rate contribution for any $n>1$. In any case, for simplicity, in our simulations we obtain bounds only for $n=1$.}

\subsection{Proof of bounds in \texorpdfstring{\cref{eq:proj_bound,eq:eigenval_bound,eq:eigenvec_bound,eq:G_function}}{Eqs. (B11) to (B13) and (B15)}}
\label{appsec:bounds_section}

\subsubsection*{\texorpdfstring{\cref{eq:proj_bound}}{Eq. (B11)}}
Let $\rho$ be a density matrix, and let $\rho ' = \frac{\Pi \rho \Pi }{\Tr[\Pi \rho \Pi]}$, where $\Pi$ is a projector. Then, 
\begin{equation}
	\begin{aligned}
		F(\rho,\rho') = \Tr[\sqrt{\sqrt{\rho}\rho'\sqrt{\rho}}]^2 = \frac{\Tr[\sqrt{\sqrt{\rho}\Pi \rho \Pi\sqrt{\rho}}]^2}{\Tr[\Pi \rho \Pi]} = \Tr[\Pi \rho \Pi],
	\end{aligned}
\end{equation}
where in the last equality we have used
\begin{equation}
	\begin{aligned}
		\Tr[\sqrt{\sqrt{\rho}\Pi \rho \Pi\sqrt{\rho}}]^2 = 	\Tr[\sqrt{\sqrt{\rho}\Pi \sqrt{\rho}\sqrt{\rho} \Pi\sqrt{\rho}}]^2 = \Tr[\sqrt{\rho}\Pi \sqrt{\rho}]^2 = \Tr[\Pi \rho \Pi]^2.
	\end{aligned}
\end{equation}
Thus, we have that
\begin{equation}
	\label{eq:proj_bound2}
	F(\rho_{\rm model}^{\mu},\rho_{\rm model}^{\prime \mu}) =  \Tr[\Pi_M \rho_{\rm model}^{\mu}\Pi_M] = \sum_{n=0}^{M} 
	p'_{\lambda_n\vert \mu} \eqqcolon  F_{\rm proj}^{\mu}.
\end{equation}

\subsubsection*{\texorpdfstring{\cref{eq:eigenval_bound}}{Eq. (B12)}}

Using Theorem 2 in Appendix A of Ref.~\cite{naharDecoyStateQuantum2022}, we have that
\begin{equation}
	\label{eq:nahar_Th2}
	\abs{p_{\lambda_n\vert \mu} - p'_{\lambda_n\vert \mu}} \leq 2 \sqrt{1-\Tr[\Pi_M \rho_{\rm model}^{\mu}\Pi_M]} = 2 \sqrt{1-F_{\rm proj}^{\mu}} \eqqcolon \epsilon_{\rm val}^{\mu}.
\end{equation}

\subsubsection*{\texorpdfstring{\cref{eq:eigenvec_bound}}{Eq. (B13)}}

Using Theorem 3 in Appendix A of Ref.~\cite{naharDecoyStateQuantum2022}{, we find that}
\begin{equation}
	\abs*{\braket*{\lambda_n^{\prime \mu}}{\lambda_n^{\mu}}}^2  \geq 1-\left(\frac{\epsilon_{\rm val}^{\mu}}{\delta_n}\right)^2 \coloneqq F_{\textrm{vec},\lambda_n}^{\mu},
\end{equation}
where $\delta_0 = p'_{\lambda_{0}\vert \mu} -  p'_{\lambda_{1}\vert \mu} - \epsilon_{\rm val}^{\mu}$ and for $n>1$,
\begin{equation}
	\delta_n = \min \{p'_{\lambda_{n-1}\vert \mu} - p'_{\lambda_{n}\vert \mu} - \epsilon_{\rm val}^{\mu},\, p'_{\lambda_{n}\vert \mu} -  p'_{\lambda_{n+1}\vert \mu} - \epsilon_{\rm val}^{\mu}\}.
\end{equation}

\subsubsection*{\texorpdfstring{\cref{eq:G_function}}{Eq. (B15)}}

We use the following result from Ref.~\cite{pereiraQuantumKey2020}. Let $\ket{u}$ and $\ket{v}$ be two pure states, and let $0 \leq  E \leq \mathbb{I}$. Then,
\begin{equation}
	\label{eq:CS_ineq}
	G_{-} \big(\ev*{ E}{v},\abs{\braket{v}{u}}^2\big) \leq \ev*{ E}{u} \leq G_{+} \big(\ev*{ E}{v},\abs{\braket{v}{u}}^2\big) 
\end{equation}
where
\begin{equation}
	G_-(y,z) =
	\begin{cases}
		g_-(y,z)  & \quad \text{if } y > 1 -z \\
		0  & \quad \text{otherwise}
	\end{cases}
	\quad \quad
	\textrm{and}
	\quad \quad
	G_+(y,z) =
	\begin{cases}
		g_+(y,z)  & \quad \text{if } y < z \\
		1  & \quad \text{otherwise}
	\end{cases}
\end{equation}
with
\begin{equation}
	g_{\pm} (y,z) = y + (1-z)(1-2y) \pm 2\sqrt{z(1-z)y(1-y)}.
\end{equation}

This result can be easily extended to mixed states. Let $\sigma$ and $\sigma'$ be any two density matrices acting on some system $S$, and let $\ket{\sigma}_{S'S}$ and $\ket{\sigma'}_{S'S}$ be purifications of these states satisfying 
\begin{equation}
	\label{eq:eq_1}
	\abs{\braket{\sigma'}{\sigma}}^2 = F(\sigma, \sigma'),
\end{equation}
which exist due to {Uhlmann}'s theorem {\cite{uhlmannTransitionProbability1976}}. Then, for any $0\leq  M \leq \mathbb{I}_S$, we have that
\begin{equation}
	\label{eq:eq_2}
	\begin{aligned}
		\textrm{Tr}[\sigma  M ] &= \ev*{\mathbb{I}_{S'} \otimes  M}{\sigma} \\
		\textrm{Tr}[\sigma'  M ] &= \ev*{\mathbb{I}_{S'} \otimes  M}{\sigma'}.
	\end{aligned}
\end{equation}
Substituting $\ket{u} \to \ket{\sigma}_{S'S}$, $\ket{v} \to \ket{\sigma'}_{S'S}$ and $ E \to \mathbb{I}_{S'} \otimes  M$ in \cref{eq:CS_ineq}, and then using \cref{eq:eq_1,eq:eq_2}, we obtain \cref{eq:G_function}, i.e.\
\begin{equation}
	\label{eq:G_function2}
	G_{-}(\textrm{Tr}[\sigma'  M], F(\sigma,\sigma')) \leq  \textrm{Tr}[\sigma  M] \leq G_{+}(\textrm{Tr}[\sigma'  M],F(\sigma,\sigma')).
\end{equation}

 	\subsection{On the dimension of the SDPs}
 	{To input the SDPs in \cref{eq:fraction_sdp_finite,eq:ph_error_sdp_finite} into a computer solver, we need to use a matrix representation for the states $\{\rho_{\rm model}^{\prime \mu}\}_{\mu}$ and their eigenvectors; for this, we need to choose a particular orthonormal basis in which to express these states, with the natural choice being $\{\ket{0},...,\ket{M}\}$. First, we find the expression
 	\begin{equation}
 		\label{eq:rho_model_fock_basis}
 		\Pi_M \rho_{\rm model}^{\mu} \Pi_M = q \, \Pi_M \rho_{\rm PR}^{\mu} \Pi_M + (1-q) \Pi_M \ketbra{\sqrt{\mu}} \Pi_M = \sum_{m,m' =0}^{M} c_{m,m'}^{(\mu)} \ketbra{m}{m'},
 	\end{equation}
 	where
 	\begin{equation}
 		\begin{cases}
 			c_{m,m}^{(\mu)} = \frac{\mu^m e^{-\mu}}{m!}, \\
 			c_{m,m'}^{(\mu)} = (1-q)\frac{\mu^{\frac{m+m'}{2} e^{-\mu}}}{\sqrt{m!m'!}} \quad m\neq m'.
 		\end{cases}
 	\end{equation}
 	Then, we numerically find the eigenvalues $\{p'_{\lambda_n\vert \mu}\}_n$ and eigenvectors $\{\ket*{\lambda_n^{\prime \mu}}\}_n$ of $\Pi_M \rho_{\rm model}^{\mu}\Pi_M $, with the latter expressed in the Fock basis
  	\begin{equation}
 	\label{eq:eigenvec_fock_basis}
 	\ket*{\lambda_n^{\prime \mu}} = \sum_{m=0}^M \sqrt{c_m^{(\lambda_n^{\mu})}} \ket{m}.
	 \end{equation}	
 	Finally, we renormalise \cref{eq:rho_model_fock_basis} to obtain the expression for $\rho_{\rm model}^{\prime \mu}$, and substitute everything into the SDPs in \cref{eq:fraction_sdp_finite,eq:ph_error_sdp_finite}.
 		

	 	Note that, while the SDP in \cref{eq:fraction_sdp_finite} does not depend on the encoding operators $\{\hat V_{0_Z},\hat V_{1_Z},\hat V_{0_X},\hat V_{1_X}\}$, the SDP in \cref{eq:ph_error_sdp_finite} does depend on the form of these operators. Typically, the output space of these operators has a larger dimension than the input space. For example, in our simulations, for simplicity, we assume that these are ideal {$Z$- and $X$-basis} BB84 operators, {whose output space consists of two modes of light and} whose action in the Fock basis is\footnote{{Note that \cref{eq:ideal_bb84_encoding_operators} represents ideal $Z$- and $X$-basis BB84 operators regardless of the physical degree of freedom used for the encoding. For time-bin encoding, the first ket would represent, say, the early time bin, and the second ket would represent the late time bin; while for polarization encoding, the first ket would represent, say, the horizontally-polarized mode, and the second ket would represent the vertically-polarized mode. \\ Also, note that it is perhaps more standard to define BB84 encoding operators as unitary, rather than just isometric, by adding an extra input mode initialized in an arbitrary pure state, say $\ket{0}$, such that the ideal operators become $\hat V_{0_Z} \ket{m} \ket{0} = \ket{m}\ket{0}$, $\hat V_{1_Z} \ket{m} \ket{0} = \ket{0}\ket{m}$, and so on. However, defining $\{\hat V_{0_Z},\hat V_{1_Z},\hat V_{0_X},\hat V_{1_X}\}$ as unitary operators with two input and two output modes throughout the manuscript would make many formulas more cumbersome and result in the analysis being less general, since it would no longer cover non-standard encoding operations in which the output encoding space is, say, one or three modes of light, rather than two.}}
	\begin{equation}
		\label{eq:ideal_bb84_encoding_operators}
		\begin{gathered}
			\hat V_{0_Z} \ket{m} = \ket{m}\ket{0}, \\ \hat V_{1_Z} \ket{m} = \ket{0}\ket{m},\\
			\hat V_{0_X} \ket{m} = \sum_k \frac{1}{\sqrt{2^m}}\sqrt{{m \choose k}} \ket{k}\ket{m-k}, \\
			\hat V_{1_X} \ket{m} = \sum_k (-1)^k \frac{1}{\sqrt{2^m}}\sqrt{{m \choose k}} \ket{k}\ket{m-k}.
		\end{gathered}
	\end{equation}
	
	The quantum states in \cref{eq:fraction_sdp_finite} can be expressed in the basis $\{\ket{0},...,\ket{M}\}$, which contains $M+1$ elements, while the states in \cref{eq:ph_error_sdp_finite} can be expressed in the basis $\{\ket{m}\ket{m'}\}_{m+m'\leq M}$, which has $\sum_{k=0}^{M} (k+1) = \frac{(M+2)(M+1)}{2}$ elements. This means that the dimension of the SDP in \cref{eq:ph_error_sdp_finite}, and therefore the time it takes to solve it, grows much more rapidly with $M$. In principle, the tightness of the bounds, and thus the resulting secret-key rate, improves as $M$ grows. However, we have found that one can only obtain very marginal key-rate improvements beyond $M = 9$, and we have chosen this value for our simulations.}

	\section{Security of the $w$-th subkey}
\label{app:security}

In the main text, we have showed that the $w$-th subkey is secure in the actual protocol if it is secure in the $w$-th alternative scenario. {Here, we give further information on the approach we use to prove the security of the $w$-th subkey in the $w$-th alternative scenario, which is based on} complementarity \cite{koashiSimpleSecurity2009}. {The first step is to} define the following virtual protocol, which is indistinguishable from the $w$-th alternative scenario from the point of view of Eve.

\vspace{10pt}

\fbox{\centerline{\begin{minipage}{0.95\textwidth}
			\vspace{5pt}
			\textbf{$w$-th virtual protocol}
			
			(1a-1c) For every round in $\mathcal{G}_w$, Alice probabilistically chooses a intensity $\mu$ and prepares $\ket{\rho_{\rm model}^{\mu}} =\sum_{n=0}^{\infty} \sqrt{p_{\lambda_n\vert \mu}} \ket{n}_{A_n} \ket*{\lambda_n^{\mu}}_B$, a purification of the state $\rho_{\rm model}^{\mu}$ given by \cref{eq:rho_model_eigendec}. Then, she measures her ancilla $A_n$, learning the value of the tag $n$. For every round in $\mathcal{G}_{\overline w}$, Alice probabilistically chooses a intensity $\mu$ and prepares $\ket{\mu}_B$. 
			
			(1d) For every round, Alice initialises an ancilla system $A_b$ ($A_{\omega}$), associated to her choice of bit (basis). Then, she applies the following encoding operation
			\begin{equation}
				\begin{alignedat}{7}
					\hat{V}_{\rm enc} \ket{0}_{A_{\omega}} \ket{0}_{A_b} \ket{\varphi}_B =& \sqrt{\frac{p_{Z_A}}{2}} &&\ket{0}_{A_{\omega}} &&(\ket{0}_{A_b} &&\hat{V}_{0_Z} \ket{\varphi}_B &&+ \ket{1}_{A_b}  &&\hat{V}_{1_Z} &&\ket{\varphi}_B) \\
					+& \sqrt{\frac{p_{X_A}}{2}} &&\ket{1}_{A_{\omega}} &&(\ket{0}_{A_b} &&\hat{V}_{0_X} \ket{\varphi}_B  &&+ \ket{1}_{A_b}  &&\hat{V}_{1_X} &&\ket{\varphi}_B),
				\end{alignedat}
			\end{equation}
			where $\ket{\varphi}_B$ refers to any state of system $B$ prepared in the previous step.
			
			(2) Bob performs a quantum nondemolition measurement\footnote{Thanks to Assumption (A8) (basis-independent detection efficiency), Bob's measurement can be decomposed into a basis-independent nondemolition measurement followed by a two-valued $Z$ or $X$ basis measurement.}, learning which rounds are detected, and announces this information.
			
			(3) For each round, Alice measures her basis ancilla $A_{\omega}$, learning her choice of basis; while Bob probabilistically chooses a basis. Both users announce the basis information for the detected rounds. The key rounds are the set of detected rounds in which Alice and Bob both chose the $Z$ basis and Alice chose the signal intensity $\mu_s$. The test rounds are the set of detected rounds in which Bob  chose the $X$ basis. 
			
			(4) For the test rounds, Alice measures her bit value ancilla $A_b$ in the computational basis, and Bob measures his photonic system in the $X$ basis. They announce and record the outcome of these measurements.
			
			(5) For the $w$-group key rounds, Alice measures her bit value ancilla $A_b$ in the $X$ basis, and Bob measures his photonic system in the $X$ basis. Let $\textbf{x}_a^w$ ($\textbf{x}_b^w$) be Alice's (Bob's) measurement results. We define the phase-error pattern of the $w$-th sifted key as $\textbf{x}_w := \textbf{x}_a^w \oplus \textbf{x}_b^w$.
			\vspace{5pt}
\end{minipage}}}
\vspace{10pt}

{To prove the security of the $w$-th subkey,} one {simply needs} to show that, before the last step of the $w$-th virtual protocol, Alice and Bob could have defined a candidate set of phase-error patterns $\mathcal{T}_w$ of size $\abs{\mathcal{T}_w} \leq 2^{H_{\rm ph}^{w,\textrm{U}}}$ such that $\textrm{Pr}[\textbf{x}_w \notin \mathcal{T}_w] \to 0$ exponentially fast as $N\to \infty$.  This implies that, if Alice and Bob apply privacy amplification to the $w$-th sifted subkey, sacrificing slightly more than $H_{\rm ph}^{w,\textrm{U}}$ bits, the final $w$-th subkey is secret \cite{koashiSimpleSecurity2009}\footnote{More precisely, if $\textrm{Pr}[\textbf{x}_w \notin \mathcal{T}_w] \leq \varepsilon$, and the users sacrifice at least $H_{\rm ph}^{w,\textrm{U}} - \log_2 \xi$ bits in PA, then the final $w$-th subkey is $\epsilon_s$-secret, with $\epsilon_s =\sqrt{2} \sqrt{\varepsilon + \xi}$. In the asymptotic regime where $N\to\infty$, we can simply assume that the parameters $\varepsilon$, $\epsilon$ and $\xi$ approach zero. \\ Also, note that the definition of the candidate set $\mathcal{T}_w$ (and the upper bound on its size) depends on the results observed in the $w$-group test rounds. However, we omit this dependence for notational simplicity.}.

In the $w$-th virtual protocol, each 
bit of the $w$-th sifted key is tagged by its value of $n$. Therefore, Alice and Bob can estimate the phase-error rate separately for the bits with different $n$. In particular, they can simply use the observed $w$-group test data to obtain an upper bound $e^{\lambda_1,\mu_s,\textrm{U}}_{\textrm{ph},w}$ on $e^{\lambda_1,\mu_s}_{\textrm{ph},w}$, the phase-error rate of the bits for which $n=1$, such that $\textrm{Pr}[e^{\lambda_1,\mu_s}_{\textrm{ph},w} > e^{\lambda_1,\mu_s,\textrm{U}}_{\textrm{ph},w}] \to 0$ as $N \to \infty$. Let $N_{\rm sift}^w$ be the size of the $w$-th sifted key, and let $q_{\lambda_1,w}$ be the fraction of its bits such that $n=1$. By assuming that these bits have at most $q_{\lambda_1,w} N_{\rm sift}^w e^{\lambda_1,\mu_s,\textrm{U}}_{\textrm{ph},w}$ phase errors, Alice and Bob can define a candidate set of phase-error patterns $\mathcal{T}_w$ of size $\abs{\mathcal{T}_w} \leq 2^{H_{\rm ph}^w}$, where
\begin{equation}
	H_{\rm ph}^w = q_{\lambda_1,w} N_{\rm sift}^w h(e^{\lambda_1,\mu_s,\textrm{U}}_{\textrm{ph},w}) + (1-q_{\lambda_1,w}) N_{\rm sift}^w,
	\label{eq:H_phw}
\end{equation}
such that $\Pr[\textbf{x}_w \notin \mathcal{T}_w] = \textrm{Pr} [e^{\lambda_1,\mu_s}_{\textrm{ph},w} > e^{\lambda_1,\mu_s,\textrm{U}}_{\textrm{ph},w}]$ approaches zero as $N \to \infty$. This implies that the $w$-th subkey key is secret if Alice and Bob sacrifice at least $H_{\rm ph}^w$ bits in the privacy amplification step.

In the actual protocol, Alice and Bob do not know which bits have a tag $n=1$, and thus cannot know the value of $q_{\lambda_1,w}$. However, they can find a lower bound $q_{\lambda_1,w}^{\rm L}$ such that $\textrm{Pr}[q_{\lambda_1,w} < q_{\lambda_1,w}^{\rm L}] \to 0$ as $N \to \infty$, and then sacrifice $H_{\rm ph}^{w,{\rm U}}$ bits in the privacy amplification step, where  $H_{\rm ph}^{w,{\rm U}}$ is computed by substituting $q_{\lambda_1,w}$ by $q_{\lambda_1,w}^{\rm L}$ in \cref{eq:H_phw}. The probability that this bound is incorrect just adds to the overall failure probability of the estimation process. Thus, the problem of proving the secrecy of the $w$-th subkey is reduced to the problem of obtaining the bounds $q_{\lambda_1,w}^{\rm L}$ and $e^{\lambda_1,\mu_s,\textrm{U}}_{\textrm{ph},w}$ using the $w$-group test data. {In {\cref{app:SDP}, we have shown} how to obtain these bounds using semidefinite programming techniques.}

Note that Alice and Bob can attempt to estimate the phase-error rate for values of $n$ other than $n=1$. In this case, the users should sacrifice
\begin{equation}
	H_{\rm ph}^{w,{\rm U}} = \sum_{n \in \mathcal{N}} q_{\lambda_n,w}^{\rm L} N_{\rm sift}^w h(e^{\lambda_n,\mu_s,\textrm{U}}_{\textrm{ph},w}) + (1- \sum_{n \in \mathcal{N}} q_{\lambda_n,w}^{\rm L}) N_{\rm sift}^w
	\label{eq:H_phw_general}
\end{equation}
bits, where $\mathcal{N}$ is the set of values of $n$ for which Alice and Bob obtain bounds $q_{\lambda_n,w}^{\rm L}$ and $e^{\lambda_n,\mu_s,\textrm{U}}_{\textrm{ph},w}$ on, respectively, $q_{\lambda_n,w}$ and $e^{\lambda_n,\mu_s}_{\textrm{ph},w}$. {As explained in {\cref{app:SDP}}, our semidefinite programming approach can be trivially modified to obtain bounds for any $n$, but in our simulations, for simplicity, we obtain bounds only for $n=1$.}

	\section{Estimation of \texorpdfstring{$q$}{q} {under the assumption $l_c = 1$}}
	\label{app:experimental_q}

	{Ref.}~\cite{kobayashiEvaluationPhase2014} argues that, when using a gain-switched laser, the phase difference $\phi_d$ between two consecutive pulses follows a Gaussian distribution, i.e.\ its PDF is
	\begin{equation}
		\label{eq:pdf_phase_difference}
		f_G(\phi_d; \hat \phi_d, \sigma),
	\end{equation}
	where the central value $\hat \phi_d$ can be assumed to be fixed throughout the experiment. {T}he standard deviation $\sigma${, on the other hand,} can be estimated by measuring the fringe visibility $V$ of the interference between consecutive pulses using an asymmetric interferometer. In particular, {it is shown in Ref.~\cite{kobayashiEvaluationPhase2014} that} $V = \vert\langle e^{i\phi_d}\rangle\vert$, where
	\begin{equation}
		\label{eq:visibility_derivation}
		\langle e^{i\phi_d}\rangle = \int_{-\infty}^{\infty} d\phi_{d} e^{i \phi_d} f_G(\phi_d; \hat \phi_d, \sigma) = \exp[-\sigma^2/2] e^{i \hat \phi_d}.
	\end{equation}
	{This means that} $V =  \exp[-\sigma^2/2]$, or equivalently
	\begin{equation}
		\label{eq:sigma_visib_rel}
		\sigma = \sqrt{2\ln(1/V)}.
	\end{equation} 
	
	{In the above description,} the phase difference $\phi_d$ follows a Gaussian distribution, {and therefore can take any value in} $\{-\infty,\infty\}$. This makes sense from an {physical} perspective: if we see the phase randomisation as a process that shifts the phase randomly from the central value $\hat \phi_d$, one can distinguish a shift by $\pi$ rad from a shift by $3\pi$ rad, the former being in principle more likely than the latter. However, note that, from the point of view of Eve, a pulse with a phase $\phi$ is indistinguishable from a pulse with a phase $\phi + 2\pi$, and so on. {Thus, from the perspective of the security proof}, the conditional PDF $f(\phi_i \vert \phi_{i-1})$ should be defined {for} $\phi_i \in [0, \, 2\pi)$ only, and to compute the probability density on some point $\phi_i$, one should sum the contributions that would fall on $\phi_i \pm 2\pi$, $\phi_i \pm 4\pi$, and so on. {Thus, we have that, if the PDF of {the physical} phase difference {between consecutive pulses} is given by \cref{eq:pdf_phase_difference}, the conditional PDF $f(\phi_i \vert \phi_{i-1})$ is given by}
	\begin{equation}
		\label{eq:wrapped_gaussian_assumption2}
		f(\phi_i \vert \phi_{i-1}) = \sum_{k=-\infty}^{\infty}  f_{\rm G} (\phi_i+2\pi k; \phi_{i-1} + \hat\phi_d, \sigma) = f_{\rm WG} (\phi_i; \phi_{i-1} + \hat\phi_d, \sigma),
	\end{equation}	
	where $f_{\rm WG}$ is the PDF of a \textit{wrapped} Gaussian distribution.
	
	{Ref.~\cite{kobayashiEvaluationPhase2014} implicitly assumes that the probability distribution of a given phase depends only on the value of the previous phase, i.e.\ $l_c=1$, and the same implicit assumption is made in Ref.~\cite{grunenfelderPerformanceSecurity2020}, indicating that this is believed to be a good approximation for many scenarios. Here, we show that, under this assumption, one can estimate the parameter $q$ needed to apply our security proof, which is defined as}
	\begin{equation}
		\label{eq:desired_bound_q}
		\frac{q}{2\pi} = \min_{\phi_{i-1},\phi_{i},\phi_{i+1}} 	f(\phi_i \vert \phi_{i-1},\phi_{i+1}),
	\end{equation}
	see \cref{eq:assumption_q}. We have that
	\begin{equation}
		\label{eq:f_cond}
		\begin{gathered}
			f(\phi_i \vert \phi_{i-1},\phi_{i+1}) = \frac{f(\phi_{i-1},\phi_i,\phi_{i+1})}{f(\phi_{i-1},\phi_{i+1})} = \frac{f(\phi_{i-1}) f(\phi_{i}\vert\phi_{i-1}) f(\phi_{i+1} \vert \phi_i,\phi_{i-1})}{f(\phi_{i-1}) f(\phi_{i+1}\vert\phi_{i-1})} \\= \frac{ f(\phi_{i}\vert\phi_{i-1}) f(\phi_{i+1} \vert \phi_i)}{f(\phi_{i+1}\vert\phi_{i-1})}
			= \frac{f_{\rm WG} (\phi_i; \phi_{i-1} + \hat\phi_d, \sigma)  f_{\rm WG} (\phi_{i+1}; \phi_{i} + \hat\phi_d, \sigma)}{f(\phi_{i+1}\vert\phi_{i-1})},
		\end{gathered}
	\end{equation}
	where in the second to last step we have used $f(\phi_{i+1} \vert \phi_i,\phi_{i-1}) = f(\phi_{i+1} \vert \phi_i)$ due to $l_c = 1$, see \cref*{eq:assumption_lc}; and in the last step we have used \cref{eq:wrapped_gaussian_assumption2}. The denominator in \cref{eq:f_cond} satisfies
	\begingroup
	\allowdisplaybreaks
	\begin{gather}
		f(\phi_{i+1}\vert\phi_{i-1}) = \int_{0}^{2\pi} d\phi_i f(\phi_{i}\vert\phi_{i-1}) f(\phi_{i+1} \vert \phi_i, \phi_{i-1}) \nonumber \\
		= \int_{0}^{2\pi} d\phi_i f(\phi_{i}\vert\phi_{i-1}) f(\phi_{i+1} \vert \phi_i) = \int_{0}^{2\pi} d\phi_i f_{\rm WG} (\phi_i; \phi_{i-1} +\hat\phi_{d}, \sigma) f_{\rm WG} (\phi_{i+1  }; \phi_{i} + \hat\phi_{d}, \sigma) \nonumber \\ 
		\stackrel{(1)}{=} \int_{0}^{2\pi} d\phi_i f_{\rm WG} (\phi_i; \phi_{i-1} +\hat\phi_{d}, \sigma) f_{\rm WG} (\phi_{i+1} -\hat\phi_{d}; \phi_{i} , \sigma) \nonumber \\ 
		\stackrel{(2)}{=} \int_{0}^{2\pi} d\phi_i f_{\rm WG} (\phi_i; \phi'_{i-1}, \sigma) f_{\rm WG} (\phi''_{i+1}; \phi_{i} , \sigma)
		\nonumber \\
		= \sum_{k=-\infty}^{\infty} \int_{0}^{2\pi} d\phi_i f_{\rm G} (\phi_i + 2 \pi k; \phi'_{i-1}, \sigma) f_{\rm WG} (\phi''_{i+1} ; \phi_{i}, \sigma) \nonumber \\
		= \sum_{k=-\infty}^{\infty} \int_{0}^{2\pi} d\phi_i f_{\rm G} (\phi_i + 2 \pi k; \phi'_{i-1}, \sigma) f_{\rm WG} (\phi''_{i+1} - 2 \pi k ; \phi_{i}, \sigma) \nonumber \nonumber \\
		\stackrel{(3)}{=} \sum_{k=-\infty}^{\infty} \int_{0}^{2\pi} d\phi_i f_{\rm G} (\phi_i + 2 \pi k; \phi'_{i-1}, \sigma) f_{\rm WG} (\phi''_{i+1} ; \phi_{i} + 2 \pi k, \sigma) \label{eq:denominator_derivations}  \\
		= \sum_{k=-\infty}^{\infty} \int_{2\pi k}^{2\pi (k+1)} d\phi_i f_{\rm G} (\phi_i; \phi'_{i-1}, \sigma) f_{\rm WG} (\phi''_{i+1} ; \phi_{i}, \sigma) \nonumber   \\ 
		=\int_{-\infty}^{\infty} d\phi_i f_{\rm G} (\phi_i; \phi'_{i-1}, \sigma) f_{\rm WG} (\phi''_{i+1} ; \phi_{i}, \sigma) \nonumber \\
		= 	\sum_{k=-\infty}^{\infty} \int_{-\infty}^{\infty} d\phi_i f_{\rm G} (\phi_i; \phi'_{i-1}, \sigma) f_{\rm G} (\phi''_{i+1}  + 2\pi k; \phi_{i}, \sigma) \nonumber \\
		= \sum_{k=-\infty}^{\infty} \int_{-\infty}^{\infty} d\phi_i f_{\rm G} (\phi_i; \phi'_{i-1}, \sigma) f_{\rm G} (\phi''_{i+1}  + 2\pi k - \phi_{i}; 0, \sigma)
		\nonumber \\
		\stackrel{(4)}{=} \sum_{k=-\infty}^{\infty} f_{\rm G} (\phi''_{i+1}  + 2\pi k; \phi'_{i-1}, \sqrt{2} \sigma) \nonumber \\
		= f_{\rm WG} (\phi''_{i+1} ; \phi'_{i-1}, \sqrt{2}\sigma), \nonumber 
	\end{gather}
	\endgroup
	where in (1) and (3) we have used $f_{\rm WG} (x; \mu, \sigma) = f_{\rm WG} (x + a ; \mu +a, \sigma)$; in (2) we have defined $\phi'_{i-1} = \phi_{i-1} + \hat\phi_{d}$ and $\phi''_{i+1} = \phi_{i+1} - \hat\phi_{d}$; and in (4) we have {used the fact that the convolution between two Gaussian PDFs $f_G(x,\mu_1,\sigma_1)$ and $f_G(x',\mu_2,\sigma_2)$ is known to be}
	\begin{equation}
		\int_{-\infty}^{\infty} d\tau f_G(\tau;\mu_2,\sigma_2) f_G (x-\tau;\mu_1,\sigma_1) = f_G\big(x;\mu_1+\mu_2,\sqrt{\sigma_1^2+\sigma_2^2}\big).
	\end{equation}
	Substituting \cref{eq:denominator_derivations} in \cref{eq:f_cond}, we have that
	\begin{equation}
		\label{eq:f_cond_final}
		f(\phi_i \vert \phi_{i-1},\phi_{i+1}) = \frac{f_{\rm WG} (\phi_i; \phi'_{i-1}, \sigma) 	f_{\rm WG} (\phi''_{i+1}; \phi_{i}, \sigma)}{f_{\rm WG} (\phi''_{i+1} ; \phi'_{i-1}, \sqrt{2}\sigma)},
	\end{equation}
	where we have again used $f_{\rm WG} (x; \mu, \sigma) = f_{\rm WG} (x + a ; \mu +a, \sigma)$ and the definition of $\phi'_{i-1}$ and $\phi''_{i+1}$.
	Finally, our desired parameter $q$ in \cref{eq:desired_bound_q} can be expressed as
	\begin{equation}
		\label{eq:expression_q}
		\frac{q}{2 \pi} = \min_{\phi_{i-1},\phi_{i},\phi_{i+1}} 	f(\phi_i \vert \phi_{i-1},\phi_{i+1}) = \min_{\phi'_{i-1},\phi_{i},\phi''_{i+1}}  \frac{f_{\rm WG} (\phi_i; \phi'_{i-1}, \sigma) 	f_{\rm WG} (\phi''_{i+1}; \phi_{i}, \sigma)}{f_{\rm WG} (\phi''_{i+1} ; \phi'_{i-1}, \sqrt{2}\sigma)}.
	\end{equation}
	
	{Ref.~\cite{grunenfelderPerformanceSecurity2020} has recently reported a fringe visibility of $V = 0.0019$ for a practical decoy-state QKD source run at a repetition rate of 5 GHz. Using this value, from \cref{eq:sigma_visib_rel}, we obtain $\sigma = 3.54003$.} Substituting {this} in \cref{eq:expression_q} and finding the exact minimum using Mathematica's \texttt{Minimize} function, we obtain
	\begin{equation}
		q = 0.992407.
	\end{equation}
	The minimum occurs when $\phi_{i} = \phi'_{i-1}\pm \pi$ and $\phi''_{i+1} = \phi'_{i-1}$.

\section{On the security analysis in Refs.~\texorpdfstring{\cite{naharQuantumKey2021,naharDecoyStateQuantum2022}}{[S2,S5]}}
\label{app:on_previous_results}

The security of decoy-state QKD with imperfect phase randomisation has {also} been {recently} investigated by Refs.~\cite{naharQuantumKey2021,naharDecoyStateQuantum2022}. These works introduced novel and insightful ideas to approach the problem that have been indispensable in the development of our security proof. However, we believe that {their} overall security analysis contains an important flaw that invalidates its application in the presence of correlations. Here, we summarise the arguments of Refs.~\cite{naharQuantumKey2021,naharDecoyStateQuantum2022} and point out what we believe to be the problem. We focus on Ref.~\cite{naharDecoyStateQuantum2022}, where the arguments are elaborated on in much more detail.

\subsubsection{Argument}

For simplicity, Ref.~\cite{naharDecoyStateQuantum2022} considers a laser source {that emits $N$ pulses} with correlated phases and a fixed intensity $\mu$, {whose state is given by}
\begin{equation}
\rho_{\rm laser}^{\mu} = \int d \phi_1 \ldots d\phi_N f(\phi_1\ldots \phi_N) \ketbra*{\sqrt{\mu} e^{i \phi_1}} \otimes \ldots \otimes \ketbra*{\sqrt{\mu} e^{i \phi_N}}.
\end{equation}
One can express the probability distribution as 
\begin{equation}
\label{eq:cond_prob_dist}
f(\phi_1 \ldots \phi_N) = f(\phi_1) f(\phi_2 \vert \phi_1) \ldots f(\phi_N \vert \phi_1 \ldots \phi_{N-1})
\end{equation}
and consider the following bound
\begin{equation}
\label{eq:q_nahar}
\frac{q}{2 \pi} \leq \min_i \min_{\phi_1 ... \phi_{i}} f(\phi_i \vert \phi_1 \ldots \phi_{i-1}).
\end{equation}
The argument of Ref.~\cite{naharDecoyStateQuantum2022} is that, instead of generating $\rho_{\rm laser}^{\mu}$, Alice could have {alternatively} generated $N$ copies of the following model state
\begin{equation}
\label{eq:rho_model_norbert}
\rho_{\rm model}^{\mu} = q \, \rho_{\rm PR}^{\mu} + (1-q) \ketbra{\sqrt{\mu}}
\end{equation}
and then applied a map $\mathcal{E}$ that consists of ``$N$ phase shifters that shift the phase of the $i$-th laser pulse by $\phi_i$ with probability [density] $\frac{f(\phi_i \vert \phi_1 \ldots \phi_{i-1}) - q/2\pi}{1-q}$''. {In doing so, one obtains} ``a correlated state from an IID state by applying a map that is correlated; the action of the $i$-th phase shifter depends on the action of all the ($i-1$) phase shifters before it''. {As a result, we have that}
\begin{equation}
\label{eq:nonexistant_operation}
\rho_{\rm laser}^{\mu} = \mathcal{E} (\rho_{\rm model}^{\mu \, \otimes N}).
\end{equation}
{Importantly, this implies} that, to prove the security, one can assume that Alice generates $\rho_{\rm model}^{\mu \, \otimes N}$ rather than $\rho_{\rm laser}^{\mu}${, since the operation $\mathcal{E}$ can be assumed to be part of Eve's attack.}

\subsubsection{Our interpretation of the argument {and its problem}}

{Given the} phase probability distribution $f(\phi_1\ldots \phi_N)${, we have that,} from the point of view of Eve, these phases could have been selected by Alice using a sequential process: she chooses $\phi_1$ according to the PDF $f(\phi_1)$, she chooses $\phi_2$ according to the conditional PDF $f(\phi_2 \vert \phi_1)$, and so on{, as indicated by \cref{eq:cond_prob_dist}}. The assumption is that $f(\phi_i \vert \phi_1 \ldots \phi_{i-1}) \geq q/2\pi$ for some $q$.

{Alternatively}, Alice could have decided the phase $\phi_i$ using the following equivalent process. {She} flips a biased coin $C_i$ that outputs $C_i=0$ with probability $q$. If $C_i=0$, Alice chooses $\phi_i^{\rm model}$ according to a uniform distribution on $[0,2 \pi)$. If $C_i = 1$, Alice chooses $\phi_i^{\rm model} = 0$. Then, Alice chooses $\phi_i^{\rm shift}$ according to the conditional PDF
\begin{equation}
\label{eq:phi_shift_norbert}
f(\phi_i^{\rm shift} \vert \phi_1 \ldots \phi_{i-1}) = \frac{f(\phi_i \vert \phi_1 \ldots \phi_{i-1}) - q/2\pi}{1-q}.
\end{equation}
Finally, Alice sets $\phi_i = \phi_i^{\rm model} + \phi_i^{\rm shift}$.

The argument of Ref.~\cite{naharDecoyStateQuantum2022} seems to be that, since $\phi_i^{\rm model}$ is chosen uniformly randomly with probability $q$, and $\phi_i^{\rm model} = 0$ with probability $1-q$, the above process is equivalent to assuming that Alice {first} generates the state {given by \cref{eq:rho_model_norbert}} 
for {each of the} rounds, and then shifts the phase of the $i$-th pulse by $\phi_i^{\rm shift}$, according to the conditional PDF in \cref{eq:phi_shift_norbert}. The {action of the} combined phase shifts $\phi_1^{\rm shift}...\phi_N^{\rm shift}$ can be represented as an overall global quantum operation $\mathcal{E}$, and thus {\cref{eq:nonexistant_operation} holds.}

		{However, we believe this argument has the following flaw. In order t}o apply the $i$-th phase shift according to the conditional PDF in \cref{eq:phi_shift_norbert}, one needs to know the previous overall phases $\phi_1 ... \phi_{i-1}$. These depend not only on the previous $i-1$ phase shifts $\phi_1^{\rm shift} ... \phi_{i-1}^{\rm shift}$, but also on the previous $i-1$ model phases $\phi_1^{\rm model} ... \phi_{i-1}^{\rm model}$. In the scenario in which Alice simply generates $\rho_{\rm model}^{\mu}$ for {each of the} rounds, {the value of $\phi_1^{\rm model} ... \phi_{i-1}^{\rm model}$} cannot be perfectly retrieved from the first $(i-1)$ copies of this state{, since two coherent states with different phases are not orthogonal, and therefore not perfectly distinguishable. This seems to imply that the operation $\mathcal{E}$ in \cref{eq:nonexistant_operation} does not exist in general.}
		
		{In contrast, the operation $\mathcal{E}_w$, which is needed in our security proof, is shown to exist in the main text. Importantly, unlike $\mathcal{E}$ in \cref{eq:nonexistant_operation}, Eve only needs to know the probability density function $f(\phi_1...\phi_N)$ to apply {$\mathcal{E}_w$}. She does not need to perform any measurement on the signals emitted by Alice.}
		
		%
		%
		
		\subsubsection{Information about the \texorpdfstring{$i$}{i}-th phase is leaked into the following pulses}
		
		In addition to the above, the idea of relating how close the $i$-th pulse is to a perfect PR-WCP by lower bounding the PDF of the $i$-th phase conditioned on the \textit{previous} phases seems to have a fundamental problem. Namely, it does not take into account that, in the presence of phase correlations, information about the $i$-th phase is leaked into the \textit{following} pulses. {To demonstrate this}, we show an example in which, using this idea, one could conclude that half of the emissions are perfect PR-WCPs, when this is clearly not the case.
		
		
		More specifically, {as discussed above,} the argument of Ref.~\cite{naharDecoyStateQuantum2022} is that, if for some round $i$ one can obtain a bound 
		\begin{equation}
			\label{eq:q_i}
			\frac{q_i}{2 \pi} \leq \min_{\phi_1 ... \phi_{i}} f(\phi_i \vert \phi_1 \ldots \phi_{i-1}),
		\end{equation}
		then one could substitute the $i$-th pulse by the generation of the state
		\begin{equation}
			\label{eq:rhomodel_i}
			\rho_{\rm model}^{\mu,(i)} = q_i \, \rho_{\rm PR}^{\mu} + (1-q_i) \ketbra{\sqrt{\mu}},
		\end{equation}
		followed by a phase shift such that the $i$-th emitted pulse ends up being identical as in the original scenario. To prove the security, it is useful to consider that the emitted state is the same for all rounds. Thus, Ref.~\cite{naharDecoyStateQuantum2022} considers instead the bound
		\begin{equation}
			\label{eq:q_alt}
			q \coloneqq \min_{i} q_i.
		\end{equation}
		and assumes that all emissions are replaced by the generation of the same IID state {given by \cref{eq:rho_model_norbert}} 
		followed by the appropriate phase shift operation for each pulse. 
		
		Now, let us consider a scenario in which Alice has a {special} source such that:
		
		\begin{enumerate}
			\item if $i$ is odd, the {emitted} pulse has a uniformly distributed phase that is independent of the phases of all previous pulses;
			\item if $i$ is even, the {emitted} pulse has a phase that is identical to that of the previous odd pulse.
		\end{enumerate}
		
		For this scenario, we have that: (1) if $i$ is odd, $q_i = 1$ and (2) if $i$ is even, $q_i = 0$. Thus, the replacement in \cref{eq:q_alt,eq:rho_model_norbert} cannot be directly used to prove the security of this {case}, since $q=0$. However, we could instead consider the security of the odd and even pulses separately. Using the argument in \cref{eq:q_i,eq:rhomodel_i}, we could assume that, in the odd rounds, Alice prepares the PR-WCP
		\begin{equation}
			\label{eq:rhomodelodd}
			\rho_{\rm model}^{\mu,\textrm{odd}} =\rho_{\rm PR}^{\mu};
		\end{equation}
		and in the even rounds, she prepares $\rho_{\rm model}^{\mu,\textrm{even}} = \ketbra*{\sqrt{\mu}}$. Then, we could simply discard all data obtained in the even rounds, and apply the standard decoy-state method to the data obtained in the odd rounds. {In doing so}, we could conclude that the secret-key rate obtainable using this source would be half of that obtainable using a source that produces perfect PR-WCPs in all rounds.
		
		However, the argument above has a {crucial} flaw: it does not take into account the fact that information about the phase of a given odd pulse $i$ is leaked into the following even pulse, and that Eve could in principle learn some of this information and use it to attack the $i$-th pulse. Thus, from Eve's point of view, the $i$-th pulse is not necessarily a PR-WCP even if its distribution is uniform when conditioned on all the \textit{previous} (but not \textit{following}) phases. This invalidates the argument in \cref{eq:q_i,eq:rhomodel_i}, which seems to be at the core of the approach in Ref.~\cite{naharDecoyStateQuantum2022}.
		
		Note that leaked information about the $i$-th phase is only useful to Eve if she can actually use it to alter the detection statistics of the $i$-th pulse. To prevent Eve from doing so, one option could be to run the protocol very slowly, such that Alice only emits the ($i+1$)-th pulse {once} Bob has finished his measurement of the $i$-th pulse. It could be possible that the security bounds derived in Ref.~\cite{naharDecoyStateQuantum2022} are correct for this scenario. However, {if the protocol is run very slowly, one does not expect that it will suffer from phase correlations, since these are mainly a problem in high-speed QKD systems.}

	\vspace{3pt}

	
%
	
%

%

\end{document}